\documentclass[conference]{IEEEtran}
\IEEEoverridecommandlockouts
\pdfoutput=1
\usepackage{subcaption}
\usepackage{xcolor} 
\usepackage{booktabs} 
\usepackage{algorithmicx}
\usepackage[ruled]{algorithm}
\usepackage{algpseudocode}
\usepackage{bold-extra}
\usepackage{setspace}
\usepackage{multirow}
\usepackage{longtable}
\usepackage[all]{xy}
\usepackage{comment}

\definecolor{stringColor}{rgb}{0.8,0.1,0.1}
\usepackage[scaled=0.88]{helvet}
\usepackage{listings}
\lstdefinelanguage{LF}{
  keywords={at, deadline, after, state, logical, physical, startup, shutdown,
  reaction, preamble, target, reactor, input, output, constructor, new,
  action, 
  actor, handler, time, main, federated, serializer, STP, timer, sec,
  secs, msec, msecs, usec, usecs},
  emph={L, trigger, 
  class, name, init, effect, instance, type, delay, 
  }, emphstyle=\itshape,
  keywordstyle=\color{black}\bfseries,
  ndkeywords={class, export, boolean, throw, implements, import, this},
  ndkeywordstyle=\color{darkgray}\bfseries,
  identifierstyle=\color{black},
  sensitive=false,
  comment=[l]{//},
  morecomment=[s]{/*}{*/},
  commentstyle=\color{purple}\ttfamily,
  stringstyle=\color{black}\ttfamily,
  morestring=[b]',
  morestring=[b]"
}
\lstdefinestyle{framed} {
  xleftmargin=15pt,
  frame=l,
  basicstyle=\scriptsize\ttfamily,
  framesep=5mm,
  fillcolor=\color{gray!10},
  rulecolor=\color{gray!10},
  numberstyle=\normalfont\tiny\color{gray!90}
}

\newenvironment{compactItemize}{\begin{list}{$\bullet$}
  {\setlength{\topsep}{1mm}\setlength{\itemsep}{0.25mm}
  \setlength{\parsep}{0.1mm}
  \setlength{\itemindent}{0mm}\setlength{\partopsep}{0mm}
  \setlength{\labelwidth}{15mm}
  \setlength{\leftmargin}{4mm}}}{\end{list}}

\newcounter{myctr}
\newenvironment{compactEnumerate}{\begin{list}{\arabic{myctr}.}
{\usecounter{myctr}
\setlength{\topsep}{0.3mm}\setlength{\itemsep}{0mm}
\setlength{\parsep}{0.3mm}
\setlength{\itemindent}{0mm}\setlength{\partopsep}{0mm}
\setlength{\labelwidth}{15mm}
\setlength{\leftmargin}{4mm}}}{\end{list}}

\usepackage{amsmath}
\usepackage{amsfonts}
\usepackage{amsthm}
\usepackage{multicol}
\usepackage[hidelinks]{hyperref}
\usepackage[inline]{enumitem}
\usepackage{xcolor} 
\usepackage{ifthen}
\usepackage{xspace}

\newcommand{\keyword}[1]{\texttt{\textbf{\small#1}}}
\newcommand{\code}[1]{\texttt{\small#1}}

\algnewcommand{\LineComment}[1]{\State \(\triangleright\) #1}

\usepackage{tikz}
\usetikzlibrary{arrows,shapes,snakes,automata,backgrounds,petri,graphs}

\newcommand{\physicalConn}{\texttt{{\raise.17ex\hbox{$\scriptstyle\mathtt{\sim}$}}>}\xspace}

\newcommand{\nn}{\mathbb{N}}
\newcommand{\ttt}{\mathbb{T}}

\newcommand{\tags}{\mathbb{G}}
\newcommand\FOREVER{$\infty$\xspace}
\newcommand\NEVER{$-\infty$\xspace}
\newcommand\NET[1][\ ]{\ensuremath{\text{NET}_{#1}}}
\newcommand\LTC[1][\ ]{\ensuremath{\text{LTC}_{#1}}}
\newcommand\EIMT[1][\ ]{\ensuremath{\text{EIMT}_{#1}}}
\newcommand\TAG[1][\ ]{\ensuremath{\text{TAG}_{#1}}}
\newcommand\NMR[1][\ ]{\ensuremath{\text{NMR}_{#1}}}

\newcommand{\lf}[0]{\textsc{Lingua Franca}\xspace}
\newcommand{\lfshort}[0]{\textsc{LF}\xspace}

\definecolor{fixmeColor}{rgb}{1.0,0.2,0.2}
\definecolor{fixmeColorJC}{rgb}{0.0,0.2,1.0}
\definecolor{fixmeColorM}{rgb}{0.2,0.8,0.2}
\definecolor{fixmeColorS}{rgb}{0.0,0.0,1.0}
\definecolor{fixmeColorCM}{rgb}{0.0,0.8,0.8}
\definecolor{fixmeColorCong}{rgb}{1.0,0.2,1.0}
\definecolor{fixmeColoriir}{rgb}{1.0,0.5,0.0}
\definecolor{fixmeColorH}{rgb}{0.1,0.8,1.0}
\definecolor{fixmeColorSL}{rgb}{0.1,0.8,1.0}

\providecommand{\e}[1]{\ensuremath{\times 10^{#1}}}

\newcommand{\sysname}{Xronos\xspace}
\newcommand{\awauto}{Autoware.Auto\xspace}
\newcommand{\rosmsg}[1]{\textsf{#1}\xspace}
\newcommand{\rosnode}[1]{\textsl{#1}\xspace}

\setitemize{noitemsep,topsep=0pt,parsep=0pt,partopsep=0pt}
\setenumerate{noitemsep,topsep=0pt,parsep=0pt,partopsep=0pt}

\usepackage{etoolbox}
\usepackage{tikz}

\usetikzlibrary{positioning}
\usetikzlibrary{backgrounds}
\usetikzlibrary{fit}
\usetikzlibrary{calc}
\usetikzlibrary{arrows.meta}
\usetikzlibrary{shapes.geometric}

\tikzstyle{action}=[draw, minimum width=3.5mm, minimum height=3.5mm, fill=yellow, regular polygon, regular polygon sides=3, inner sep=0]
\tikzstyle{pevent}=[draw, circle, minimum width=1.5mm, inner sep=0, fill=black]
\tikzstyle{shortarrow}=[-Latex,shorten >=1mm,shorten <=1mm]
\tikzstyle{computation}=[draw, minimum width=8mm, minimum height=3mm, fill=lightgray, inner sep=0,anchor=south west]

\newcommand{\controller}[0]{\code{MPC Controller}\xspace}
\newcommand{\lgsvl}[0]{\code{LGSVL}\xspace}
\newcommand{\planner}[0]{\code{Behavior Planner}\xspace}
\newcommand{\rti}[0]{\code{RTI}\xspace}

\newcommand{\zerodisplayskips}{%
  \setlength{\abovedisplayskip}{2pt}%
  \setlength{\belowdisplayskip}{2pt}%
  \setlength{\abovedisplayshortskip}{2pt}%
  \setlength{\belowdisplayshortskip}{2pt}}
\appto{\normalsize}{\zerodisplayskips}
\appto{\small}{\zerodisplayskips}
\appto{\footnotesize}{\zerodisplayskips}

\begin{document}
\counterwithout{lstlisting}{section} 
\title{\sysname: Predictable Coordination for Safety-Critical Distributed Embedded Systems}

\makeatletter
\newcommand{\linebreakand}{%
  \end{@IEEEauthorhalign}
  \hfill\mbox{}\par
  \mbox{}\hfill\begin{@IEEEauthorhalign}
}
\makeatother

\author{
    \IEEEauthorblockN{Soroush Bateni\IEEEauthorrefmark{1},
        Marten Lohstroh\IEEEauthorrefmark{1},
        Hou Seng Wong\IEEEauthorrefmark{1},
        Rohan Tabish\IEEEauthorrefmark{2},
        Hokeun Kim\IEEEauthorrefmark{3},\linebreakand
        Shaokai Lin\IEEEauthorrefmark{1},
        Christian Menard\IEEEauthorrefmark{4},
        Cong Liu\IEEEauthorrefmark{5},
        Edward A. Lee\IEEEauthorrefmark{1}}\linebreakand
    \IEEEauthorblockA{\IEEEauthorrefmark{1}EECS Department, UC Berkeley, USA \linebreakand
        \{soroush, marten, housengw, shaokai, eal\}@berkeley.edu}
    \IEEEauthorblockA{\IEEEauthorrefmark{2}Department of Computer Science, UIUC, USA \linebreakand
        rtabish@illinois.edu}
    \IEEEauthorblockA{\IEEEauthorrefmark{3}Department of Electronic Engineering, Hanyang University, Korea \linebreakand
        hokeun@hanyang.ac.kr}
    \IEEEauthorblockA{\IEEEauthorrefmark{4}Chair for Compiler
    Construction, TU Dresden, Germany \linebreakand
        christian.menard@tu-dresden.de}
    \IEEEauthorblockA{\IEEEauthorrefmark{5}Department of Computer Science, UT Dallas,
    USA \linebreakand
        cong@utdallas.edu}
}

\date{}
\maketitle
\thispagestyle{plain}
\pagestyle{plain}
\begin{abstract}
  Asynchronous frameworks for distributed embedded systems, like ROS and
  MQTT, are increasingly used in safety-critical applications such as
  autonomous driving,
  where the cost of unintended behavior is high. The coordination mechanism
  between the components in these frameworks, however, gives rise
  to nondeterminism, where factors such as communication timing can lead to arbitrary
  ordering in the handling of messages. In this paper, we demonstrate
  the significance of this problem in an open-source full-stack autonomous driving
  software, \awauto 1.0, which relies on ROS 2. We give an alternative:
  \sysname, an open-source framework for distributed embedded systems that has a
  novel coordination strategy with predictable properties under clearly stated assumptions.
  If these assumptions are violated, \sysname provides for application-specific
  fault handlers to be invoked.
  We port \awauto to \sysname and show that it avoids the identified problems
  with manageable cost in end-to-end latency. Furthermore, we
  compare the maximum throughput of \sysname to ROS 2 and MQTT using
  microbenchmarks under different settings, including on three different 
  hardware configurations, and find that it can match or exceed those frameworks
  in terms of throughput.
\end{abstract}


\section{Introduction}\label{sec:introduction}

Frameworks such as the Robot Operating System (ROS)~\cite{quigley2009ros} and
MQTT~\cite{stanford1999mq,light2017mosquitto} are widely used in
safety-critical, concurrent, and often distributed applications such as
autonomous driving and industrial automation 
\cite{Kato:2018:Autoware, Malavolta:2020:HowToArchitectARobot, mishra2020use}.
These frameworks are convenient, modular, and their underlying asynchronous
coordination mechanism, called publish-subscribe (pub-sub), is easy to use and
not prone to deadlocks. This paper empirically shows that pub-sub is ill-suited
for such applications and offers an alternative: an open-source middleware
called \sysname built on top of an open-source coordination language, \lf
(\lfshort)~\cite{LohstrohEtAl:21:Towards}. \lfshort, based on the reactor
model~\cite{Lohstroh:EECS-2020-235}, is a polyglot coordination language that
borrows the best semantic features of established models of computation, such as
actors~\cite{Agha:97:Actors}, logical execution time (LET)~\cite{Kirsch:12:LET}, synchronous reactive
languages~\cite{Benveniste:91:Synchronous}, and discrete event
systems~\cite{LeeEtAl:7:DiscreteEvents} such as DEVS~\cite{Zeigler:1997:DEVS}
and SystemC~\cite{Liao:97:Scenic}. \lfshort furthers the state of the art by
making time a first-class citizen in the programming model and by enabling deterministic
interactions between multiple physical and logical timelines~\cite{LohstrohEtAl:21:Towards}.

The \sysname runtime system is implemented in C to ensure efficiency. \sysname applications are modular,
just like ROS and MQTT, allowing independent
processes to be deployed to distributed embedded hardware. This property is
crucial for complex distributed robotics applications such as autonomous
driving. \sysname enables predictable coordination between software components
using explicit temporal semantics that is realistic about the inability to
perfectly control timing and to perfectly synchronize clocks. It is also
realistic about the unavoidability of faults. Things \emph{will} go wrong, so
the emphasis in \sysname is on \emph{detecting} timing faults and enabling
application logic to react to them. \sysname does not require real-time network
services such as TSN~\cite{LoBelloEtAl:19:TSN} nor real-time operating system
services, but it can benefit from them to reduce the frequency of faults, to
reduce latencies, or to increase throughput.

We start with an open-source self-driving car application called \awauto and
identify subtle problems that arise due to the use of ROS's pub-sub
coordination fabric. We then port \awauto to \sysname, demonstrating that it
is equally convenient, modular, and easy to use, and that ROS apps are not hard
to convert. We show that some of the identified problems disappear and that
previously undetectable faults caused by violations of timing requirements,
become detectable, enabling the addition of application logic for dealing with
such faults. 

\sysname uses logical time to provide deterministic
concurrency, achieving better repeatability than pub-sub. \sysname aligns its logical
timeline with measurements of physical time to facilitate real-time interactions
with sensors and actuators. It supports asynchronous injection of external
events, and once a logical time has been assigned to such events, their handling
is deterministic. We show that this determinism does not incur a significant performance
cost on three test platforms, a PC, an NVIDIA Jetson AGX Xavier, and a
heterogeneous two-node distributed embedded system.


For distributed embedded systems specifically, how to deal with faults and degradations is
application-dependent. If communication latency increases, for example, some
applications will require timely reactions even with incorrect or inconsistent
data, whereas for other applications the correctness of responses is more important than their timeliness.
For example, an emergency braking system may prefer to apply the brakes
with incomplete sensor data, whereas a car may prefer to delay entering an
intersection when sensor data is incomplete or inconsistent.  

\begin{figure*}[t]
    \centering
    \includegraphics[width=0.9\linewidth]{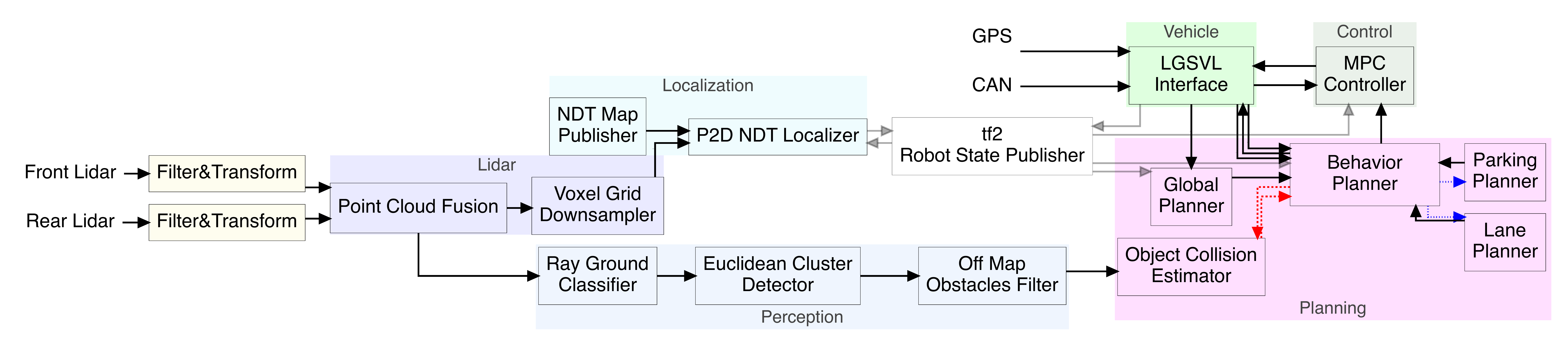}
    \caption{Architecture of \awauto 1.0 capable of Autonomous Vehicle Parking (AVP).}
    \label{fig:autoware-avp}
\end{figure*}

Brewer's CAP theorem~\cite{Brewer:12:CAP} shows that no system can have both
consistency (here, conformance with the application specifications) and
availability (here, timely reactions) when the network is partitioned. This paper
proposes two distributed coordination mechanisms, both implemented in \sysname.
One emphasizes availability over consistency, and the other emphasizes
consistency over availability when the network fails or degrades. The first mechanism,
our \textit{decentralized coordinator}, is an extension of
PTIDES~\cite{Zhao:07:PTIDES}, a real-time technique also implemented in Google
Spanner~\cite{CorbettEtAl:12:Spanner}, a globally distributed database. 
This coordinator is also influenced by Lamport~\cite{Lamport:84:TimeStamps}, and Chandy and
Misra~\cite{ChandyMisra:79:DDE,Chandy:86:DDE}.

The
second mechanism, our \textit{centralized coordinator}, is an extension of the
High-Level Architecture 
(HLA)~\cite{dahmann1997department}, a distributed discrete-event simulation
standard.  Our extensions adapt HLA's techniques to make them usable not just
for simulation, but also for
distributed real-time deployment where unpredictable asynchronous events are
injected via sensors.

The decentralized coordinator ensures software components
continue to react even if the communication latencies increase, whereas the
centralized coordinator ensures the software components behave as specified
even if their inputs are delayed.
Hence, the decentralized coordinator emphasizes availability over consistency,
whereas the centralized coordinator emphasizes consistency over availability
when the network degrades.

The rest of this paper is organized as follows: We give background
on state-of-the-art frameworks for distributed embedded systems, \awauto, and \lfshort in
Sec.~\ref{sec:background}. We demonstrate three specific issues in \awauto
with empirical evidence in Sec.~\ref{sec:motivation}. We explain the design of
\sysname and detail our extensions to \lfshort in Sec.~\ref{sec:design} and
offer implementation details in Sec.~\ref{sec:implementation}.
Finally, we evaluate \sysname on microbenchmarks and \awauto in
Sec.~\ref{sec:evaluation}, give an overview of related work in
Sec.~\ref{sec:related}, and conclude in Sec.~\ref{sec:conclusion}.

\section{Background}\label{sec:background}
\subsection{Pub-Sub Frameworks for Distributed Embedded Systems}
\noindent\textbf{ROS}\footnote{\href{https://www.ros.org/}{https://www.ros.org/}}
is a collection of tools and libraries that facilitate the development of
robotics applications. Developers can package software modules as independent
entities called nodes. Each node will live in its own separate process on a host
operating system. Nodes can co-exist on a single machine or be distributed across multiple
and communicate over the network.
Irrespective of physical location, nodes communicate using a pub-sub
model, in which publishers advertise topics and subscribers bind a specific
callback function to a topic. ROS 2, which we use in this paper, utilizes a
DDS-compliant communication framework~\cite{Thomas2014} to implement the pub-sub mechanism. We identify this underlying pub-sub as a
potential source of concurrency errors that affect the application logic and are
hard to detect and remedy (see Sec.~\ref{sec:motivation}).
\textbf{MQTT}~\cite{standard2019mqtt} is a TCP/IP-based pub-sub
protocol with similar properties that is widely used in Internet of Things applications.

\subsection{Autoware}
Autoware is an open-source software for autonomous vehicles based on ROS.
\awauto\footnote{\href{https://www.autoware.org/autoware-auto/}{https://www.autoware.org/autoware-auto/}}
is the current generation of Autoware, and is a successor to the previous version
called Autoware.AI. \awauto is based on ROS 2 and features a modular design,
consisting of a variety of nodes that are capable of perception, localization,
planning, and control.

Fig.~\ref{fig:autoware-avp}\footnote{Produced based on output from the
\code{rqt-graph} ROS package.} shows the collection of nodes that are present
in \awauto release 1.0 (black edges portray pub-sub ``topics,'' blue edges
delineate actions, and red edges illustrate services). These nodes together form
a pipeline that is capable of ``autonomous valet parking,'' where the vehicle
can go and park autonomously anywhere in a parking lot.


\awauto's modular design enables execution on a variety of platforms in a
distributed manner. For communication among distributed nodes, \awauto mainly
uses pub-sub messages and asynchronous callbacks. The distributed
design of \awauto allows concurrent execution of various tasks on dedicated
hardware; however, this also makes it challenging to achieve reproducible system
behavior.


\subsection{\lf}\label{subsec:lf-background} 
\lf (\lfshort) is an open-source polyglot coordination
language that emphasizes deterministic interaction between concurrent reactive
components called \textbf{reactors}~\cite{LohstrohEtAl:21:Towards,Lohstroh:2019:CyPhy,Lohstroh:EECS-2020-235}. 
We have chosen \lfshort as the basis for our work because of its deterministic semantics.
%
Events in \lfshort are tagged and can be sent from a port of one reactor to the port
of another.  Every event occurs at
a logical tag $g$ drawn from a totally-ordered set $\tags$ and every reactor
processes events in tag order. Each \textbf{tag} is a pair of numbers, a timestamp $t
\in \ttt$ and a microstep $m \in \nn$ (to realize superdense
time~\cite{MannaPnueli:93:Verifying}). A timestamp $t \in \ttt$ represents a
measure of time.

The functionality of a reactor is encoded by its
\textbf{reactions}, which are triggered by events and can produce new
ones. \lfshort is \emph{polyglot} in the sense that reactions are written
in one of several target languages (currently C, C++, Python, TypeScript, or Rust),
and an \lfshort program is compiled into a program in that target language.


\begin{figure}[t]
    \centering
\begin{lstlisting}[style=framed,language=LF,escapechar=|]
  target L;
  reactor class { |\label{ln:definition}|
      input name:type    |\label{ln:input}|
      output name:type    |\label{ln:output}|
      state name:type(init)   |\label{ln:state}|
      timer name(offset, period) |\label{ln:timer}|
      logical action(offset) name:type |\label{ln:laction}|
      physical action name:type |\label{ln:paction}|
      ...
      reaction(trigger, ...) source, ... -> effect, ... |\label{ln:reaction}|
      {= 
          ... code in language L ...
      =}
      ... more reactions ...
  }
  ...
  main reactor { |\label{ln:main}|
      instance = new class() |\label{ln:instance}|
      ...
      instance.name -> instance.name after delay |\label{ln:after}|
      ...
  }
\end{lstlisting}
 \caption{Structure of \lfshort programs in target language $L$}
 \label{lst:example}
\end{figure}
The \lfshort syntax is mostly concerned with the definition of reactor classes,
their instantiation, and their composition.
Lst.~\ref{lst:example} sketches an \lfshort program written in a fictional
target language \code{L} (keywords are in bold, metavariables are italicized),
showing a reactor definition (Ln.~\ref{ln:definition}) with various kinds of
class members, and a \textbf{main reactor} (Ln.~\ref{ln:main}) showing syntax
for instantiating and composing reactor instances. Inputs and outputs allow one
reactor instance to be connected to another (Ln.~\ref{ln:after}). 
The \code{->} operator on
Ln.~\ref{ln:after}, which creates a \textbf{logical connection} between an
upstream port and a downstream one, has an optional \textbf{after} clause that
adds a \textit{delay} offset to the tag of relayed events. Alternatively, the
\physicalConn operator can be used to specify a \textbf{physical connection},
which discards event tags and replaces them with a measurement of physical time.
State
variables are local to a reactor instance, and can be read or written to by any
reaction of that instance. 
Timers generate periodic
events, while actions provide sporadic events that are scheduled dynamically
through a runtime API.
Asynchronous external events can be injected into the system with the use of a \textbf{physical action}. 
The events of a physical action are assigned a tag based on a
measurement of physical time.
 
Time is a first-class data type in \lfshort, and application code has access to both the logical
clock (tags) and the physical clock on the local platform.
By default, logical time ``chases'' physical time
in a program execution, so that events with a logical time $t$ occur close to (but never before) physical time $T = t$.
Reactions may have \textbf{deadlines}
that guide an earliest deadline first (EDF) scheduler.
Application code can provide a \textbf{deadline handler},
fault-handling code to be invoked instead of a regular reaction when a deadline is violated.

Execution of an \lfshort program must ensure that each reactor reacts to
events (input messages, timer events, and actions) in tag order. 
If two events
have the same tag, they are \textbf{logically simultaneous}, and no reaction at
that tag can observe one event as present and the other as absent. Moreover,
along any communication channel between reactors and for any timer or action,
there can be at most one event with any given tag $g$. 

Our main contribution in this paper is to extend these properties across a distributed system, thereby preserving determinism.
We extend \lfshort to allow top-level reactors to become 
independent processes (federates), deployable to remote machines.
This extension includes handling of a multiplicity of physical clocks and
permitting federates to independently advance their local tags.
Our extensions enable all existing valid \lfshort programs that use the C target
(where reactions are written in C/C++) or the Python target to be transparently converted to federated programs.
It also supports specification of fault handling code that is invoked when communication
latencies increase enough to make it impossible to enforce the consistency requirements specified in the program while
keeping the program responsive.


\section{Motivational Case Study: \awauto}\label{sec:motivation}
 
\begin{figure*}[!ht]
    \centering
    \begin{subfigure}[b]{0.30\linewidth}
        \hspace{2em}
        \includegraphics[width=0.65\linewidth]{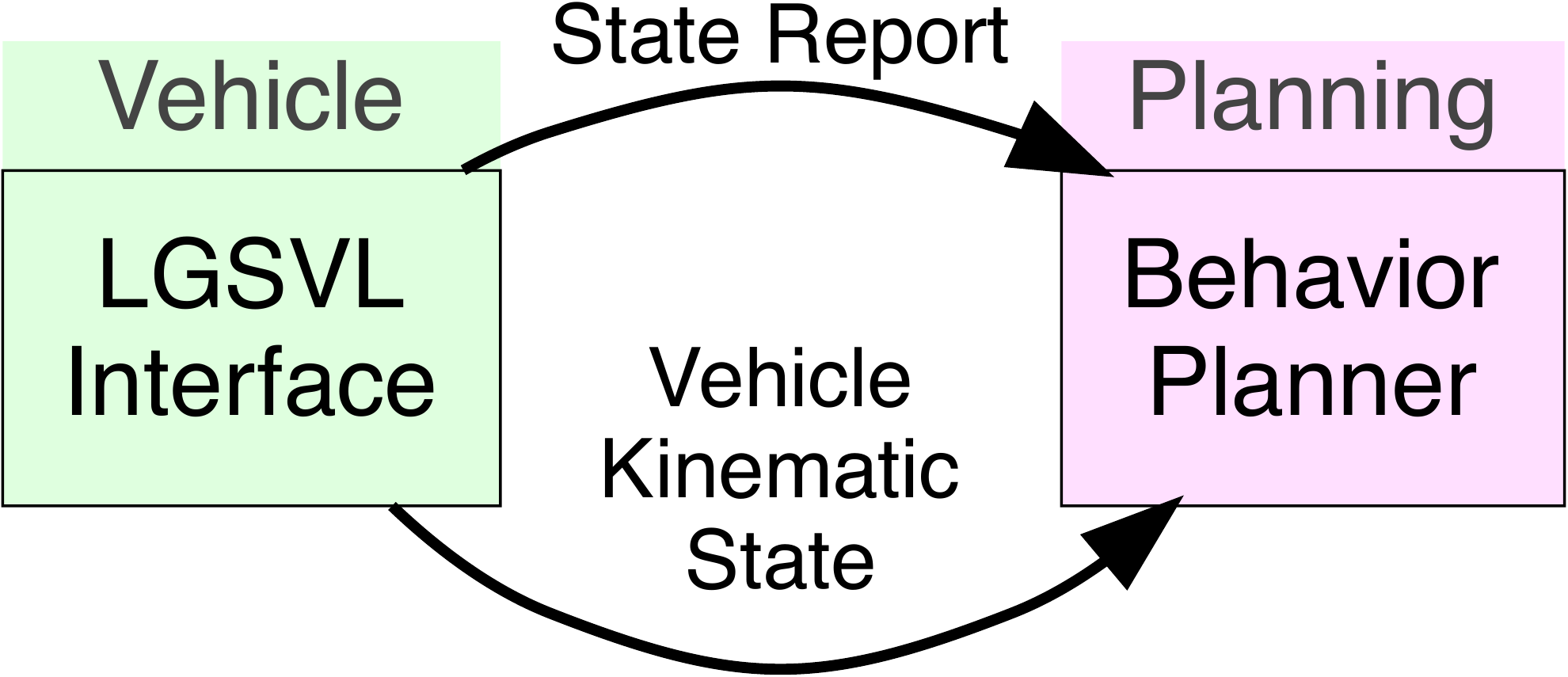}
        \vspace{1em}
        \caption{Communication pattern.}
        \label{subfig:nd-n2n}
    \end{subfigure}
    \begin{subfigure}[b]{0.34\linewidth}
        \includegraphics[width=.9\linewidth]{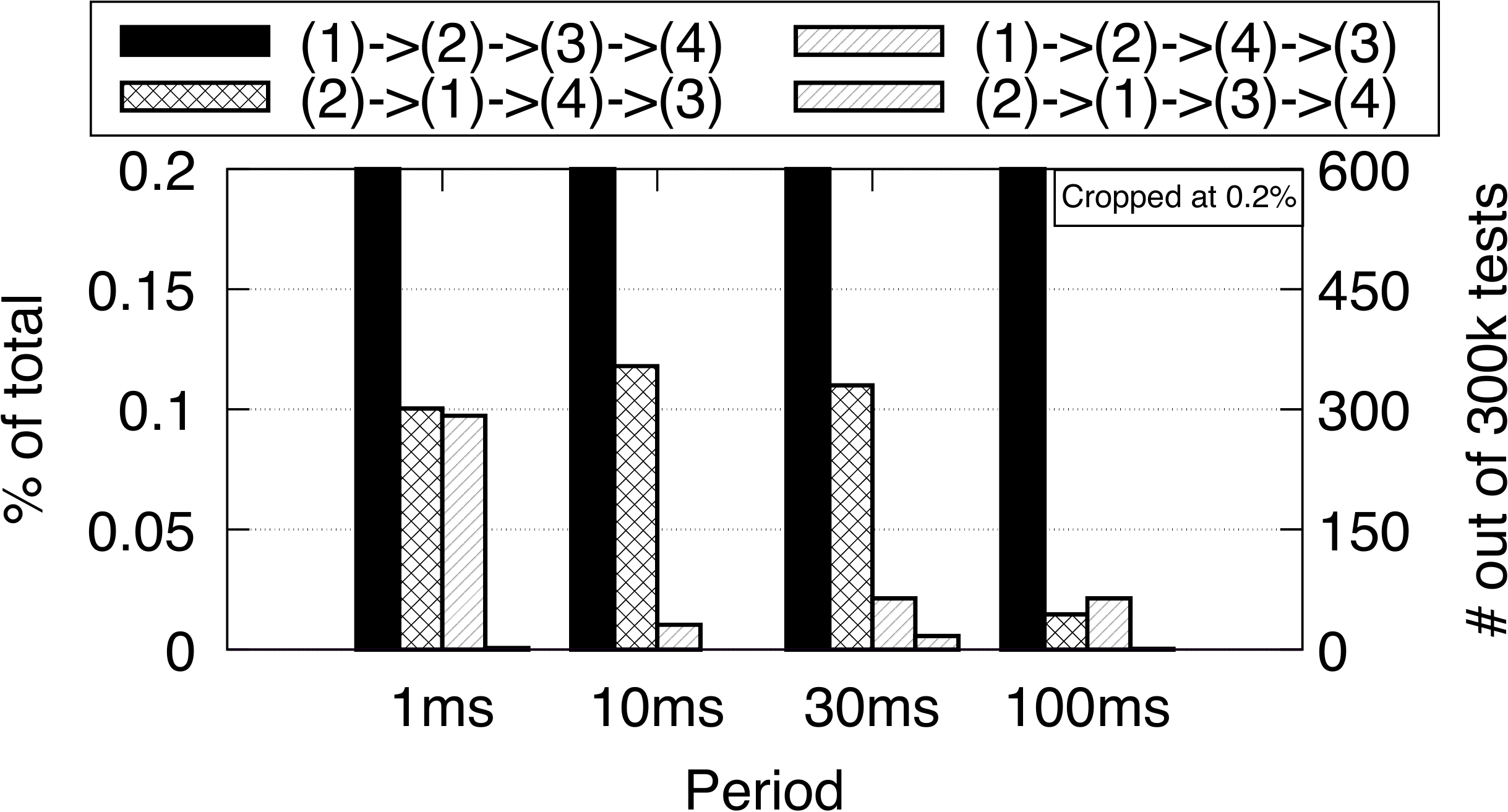}
        \caption{Periodic tests with wide gaps.}
        \label{subfig:nd-n2n-above1ms}
    \end{subfigure}
    \begin{subfigure}[b]{.34\linewidth}
        \includegraphics[width=.9\linewidth]{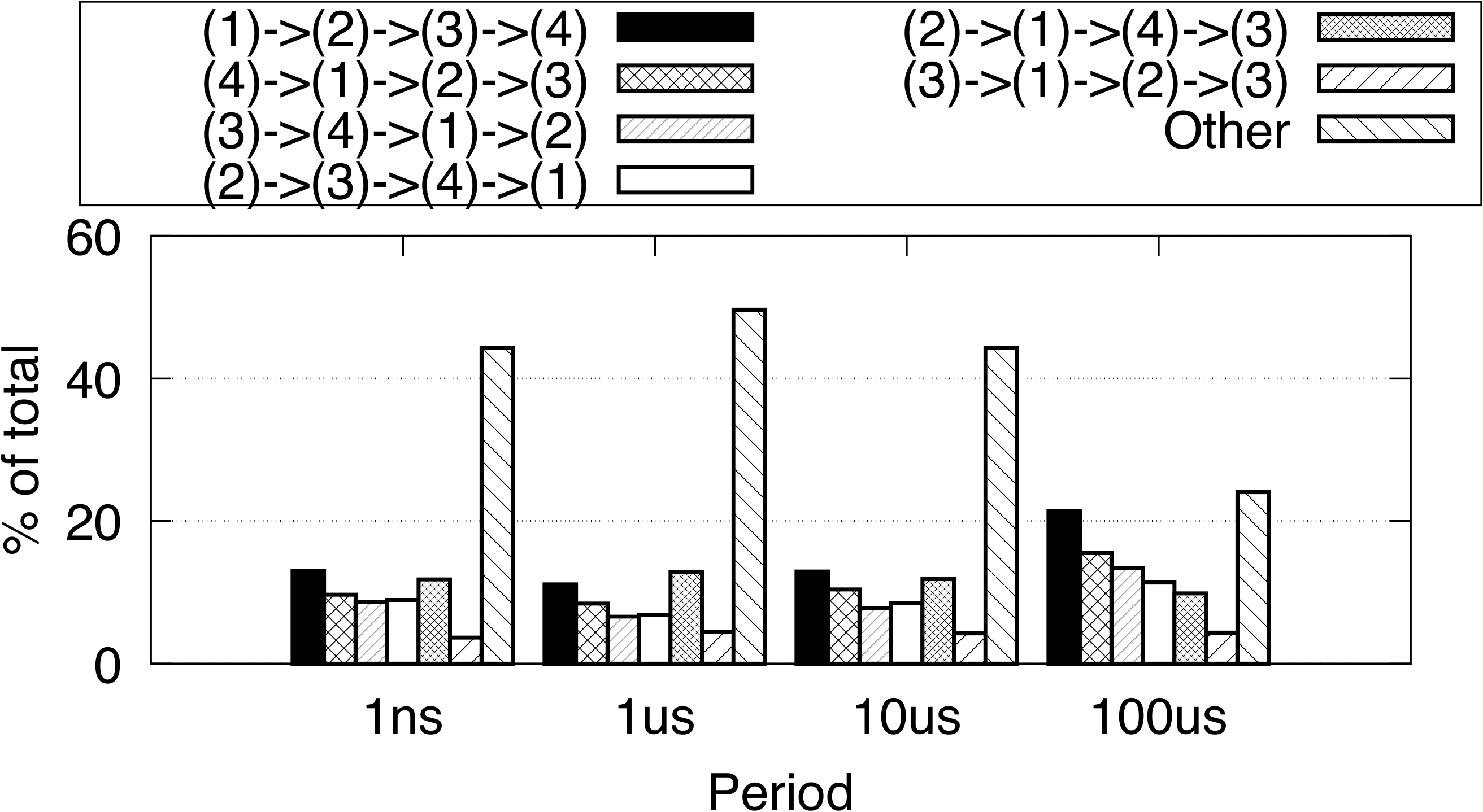}
        \caption{Periodic tests with small gaps.}
        \label{subfig:nd-n2n-below1ms}
    \end{subfigure}
    \caption{Node-to-node inconsistency in \awauto.}
    \vspace{-3mm}
\end{figure*}

To demonstrate the degree to which pub-sub-based communication
methods can undermine confidence in safety-critical applications, we consider
\awauto, a full-stack autonomous driving framework that uses ROS. 

\subsection{Node-to-Node Inconsistency}\label{subsubsec:n2n}
Consider the subsystem of \awauto depicted in Fig.~\ref{subfig:nd-n2n}. The
\rosmsg{State Report} topic includes state information about the vehicle,
including the current gear. Consecutive messages published by the \rosnode{LGSVL
Interface} (the interface to the vehicle, a simulator in our case) on this topic
are always delivered in the order they were sent. For example, if the gear is
reported as ``drive'' in one message and as ``reverse'' in the next, then the
\rosnode{Behavior Planner} will see that the vehicle has been in the ``drive''
gear and subsequently switched to ``reverse.'' This order is crucial for the
\rosnode{Behavior Planner} to correctly keep track of the current state of the
vehicle and to make accurate safety-critical decisions in general.

Notice that the \rosnode{LGSVL Interface} node also publishes messages on the
\rosmsg{Vehicle Kinematic State} topic, which includes information such as the
current velocity (positive for forward movement and negative for reverse) and the wheel angle of the vehicle, among other details.
However, messages published on these distinct topics may be delivered in
arbitrary order; the pub-sub semantics provides no built-in means to
order messages across the two independent message flows.

The \rosnode{Behavior Planner} simply stores the gear information upon receiving
the \rosmsg{State Report}. Upon receiving the \rosmsg{Vehicle Kinematic State},
it then uses the stored gear data to calculate and publish a new trajectory for
the vehicle. The output of the \rosnode{Behavior Planner} depends on the order in which
these messages arrive. Imagine the following test scenario, where the
\rosnode{LGSVL Interface} publishes messages in the following order: $(1)$
``drive'' on \rosmsg{State Report}; $(2)$ \rosmsg{Vehicle Kinematic State} with
a positive velocity; $(3)$ ``reverse'' on \rosmsg{State Report}; and $(4)$
\rosmsg{Vehicle Kinematic State} with a negative velocity.

The \rosnode{Behavior Planner} can observe the produced message sequence in any
permutation that preserves the ordering of messages within the same topic, which
we confirmed empirically. Fig.~\ref{subfig:nd-n2n-above1ms} shows the incidence
of observed message sequence permutations as a percentage of the total number of
complete message sequences published. In our experiments, we varied the rate at
which the \rosnode{LGSVL Interface} node produces sequences of
$(1)\rightarrow(2)\rightarrow(3)\rightarrow(4)$. The period after which the
sequence repeats is shown on the $y$-axis. We chose periods that correspond to
the frequency at which sensors typically produce data in the real world. We ran
each test 3\e{5} times on an Ubuntu PC with an AMD Ryzen 5800X CPU and 32GB of
DDR4 memory running ROS 2 Foxy.\footnote{The DDS runtime used here is Eclipse
Cyclone. Other DDS implementations could give more repeatable behavior. However,
our goal is not just repeatable behavior, but a deterministic semantics.}

One would normally expect the message sequence observed at the \rosnode{Behavior
Planner} to be the same as it was published by the \rosnode{LGSVL Interface}. It
would only be reasonable to expect the vehicle interface and the planner to
agree on the state trajectory of the vehicle. However, disagreement about the
order of events can lead to inconsistent views on the state of the system, and
can make the outcome of tests misleading. 

Consider the sequence $(1)\rightarrow(2)\rightarrow(4)\rightarrow(3)$, which can
be observed under minimal stress on the ROS 2 infrastructure. In this
permutation of the published message sequence, the vehicle is in the ``drive''
gear $(1)$ and reports a positive velocity $(2)$. However, before the
state of gear can be updated to ``reverse'' $(3)$, the \rosnode{Behavior
Planner} will receive a negative velocity state $(4)$. This state has to be
treated by the application developer either as a genuine fault condition, or be
dismissed as an inconsequential inconsistency introduced by the ROS framework.
In the former case, tests will transiently fail. In the latter case, tests will
not fail, but this comes at the cost of not being able to detect a dangerous
system state that might occur while the system is operating.

Even if the dangerous state is treated as a legitimate error, without any stress
on ROS, a test that checks for this condition will pass 99.8\% of the time.
This can lull the application designer into a false sense of confidence in the
safety of the application. Of course, infrequent errors due to message sequence
permutations can be made more frequent by applying stress on the ROS 2
framework. We also ran our test scenario under stress, depicted in
Fig.~\ref{subfig:nd-n2n-below1ms} with periods smaller than 1 millisecond. This
test puts pressure on the underlying message delivery infrastructure by flooding
it with messages.\footnote{This method is somewhat inspired by chaos engineering
techniques~\cite{rosenthal2020chaos}.} 
Under stress, the ROS framework buckles, and delivers messages across topics in
a multitude of seemingly unpredictable permutations. While the expected
observation order is still in the majority, the error rate is two orders of
magnitude higher than in Fig.~\ref{subfig:nd-n2n-above1ms}. We also observe
messages being dropped due to limited buffering capacity in the DDS framework,
causing odd message patterns such as
$(3)\rightarrow(1)\rightarrow(2)\rightarrow(3)$.

One solution to the node-to-node inconsistency problem is
to consolidate topics. But this severely impairs the modularity of the
application and the reusability of components. In the case of \awauto, the
\rosmsg{State Report} gets published in a superclass of the \rosnode{LGSVL
Interface}, which publishes the \rosmsg{Vehicle Kinematic State}. The two topics
have different uses, and their data are sourced from different subsystems
entirely (the CAN bus and GPS/GNSS, respectively). Another solution would be to
explicitly tag each message (e.g., using sequence numbers) within the
\rosnode{LGSVL Interface} and do manual alignment. This approach, however,
carries significant overhead and is error-prone. We propose to solve the
problem in the communication layer, where it can be done more robustly and
efficiently while keeping the application simple and modular.

\subsection{Multi-node Inconsistency} \label{subsec:multi-node} The asynchrony of
pub-sub communication poses further challenges. Consider another
subsystem of \awauto, depicted in Fig.~\ref{subfig:nd-t1}. In it, the
\rosnode{MPC Controller} produces a \rosmsg{Vehicle Command} (not depicted) on the basis
of several inputs, including a \rosmsg{Trajectory} and a \rosmsg{Vehicle Kinematic
State}. Upon receiving any input, the \rosnode{MPC Controller} will try to
recompute the \rosmsg{Vehicle Command}. The problem of multi-node inconsistency
concerns the alignment of the observed inputs. To compute an accurate
\rosmsg{Vehicle Command}, all inputs must relate to the same frame of
reference~\cite{foote2013tf}, but due to the asynchronous communication between nodes, they typically do not.

In our experimental setup, the \rosnode{Behavior Planner}  takes an average of
75 milliseconds to compute a new \rosmsg{Trajectory} upon receiving a
\rosmsg{Vehicle Kinematic State}. The \rosmsg{Vehicle Kinematic State} is
produced with a frequency of 30 Hz, resulting, on average, in 2.48 messages
being received at the \rosnode{MPC Controller} for each \rosmsg{Trajectory}. As
a consequence, the latest \rosmsg{Vehicle Kinematic State} is always newer than
the latest \rosmsg{Trajectory}. Therefore, the \rosnode{MPC Controller} attempts
to transform the last received \rosmsg{Vehicle Kinematic State} to the frame of
reference of the last received \rosmsg{Trajectory}, a costly operation that could be eliminated if proper alignment
was ensured by an underlying framework.

One could argue that transforming inputs to match the desired frame of reference
is a satisfactory solution~\cite{foote2013tf}. While this may be true for
variables such as velocity, which only marginally change over the span of a few
seconds, it is not the case for variables that encode \emph{modes} and thus
are subject to discrete changes. The presence of these kinds of variables raises
potential for ``mode confusion''~\cite{luttgen1999analyzing} in the control
system, a situation where the system executes invalid logic with respect to its
current mode. We found that such scenario can also occur in \awauto. 
\begin{figure}[t]
    \centering
    \begin{subfigure}[b]{0.60\linewidth}
        \includegraphics[width=\linewidth]{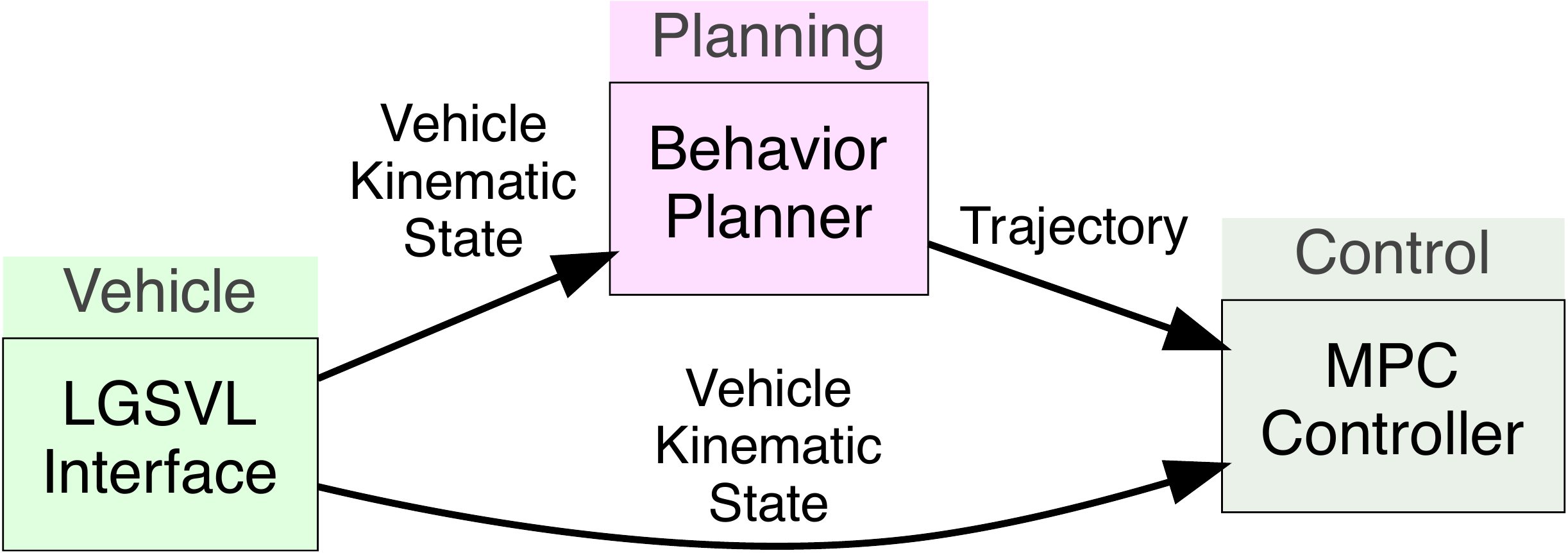}
        \caption{}
        \label{subfig:nd-t1}
    \end{subfigure}
    \begin{subfigure}[b]{0.60\linewidth}
        \includegraphics[width=\linewidth]{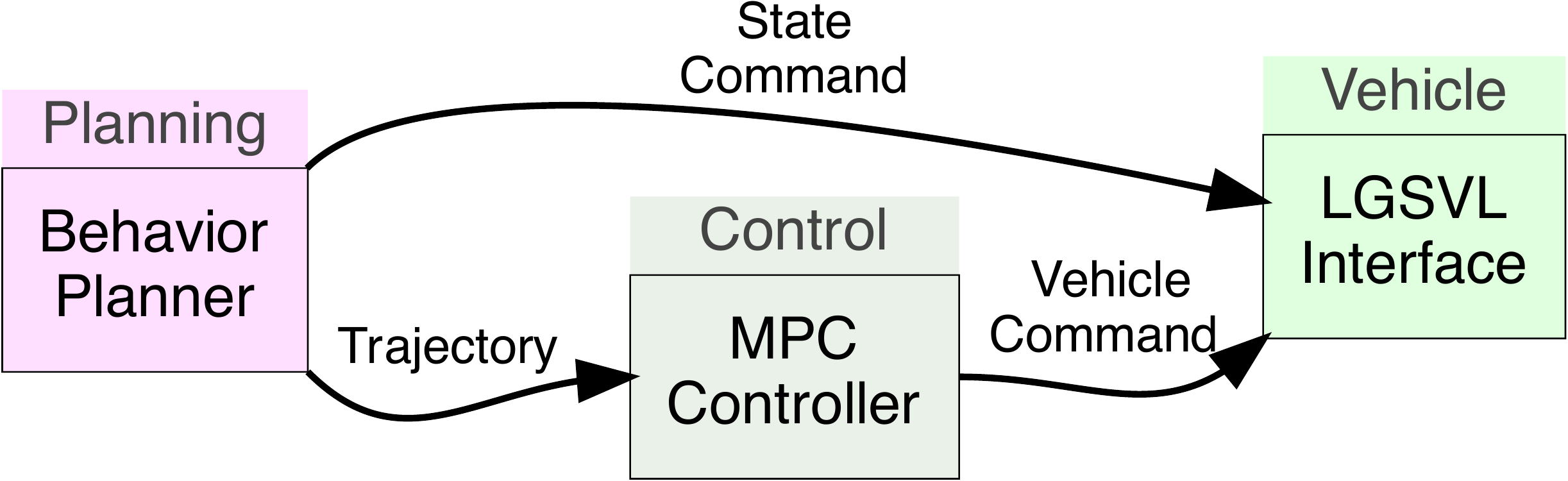}
        \caption{}
        \label{subfig:nd-t2}
    \end{subfigure}
    \caption{Multi-node inconsistency in \awauto.}
    \vspace{-3mm}
\end{figure}

Consider the subsystem shown in Fig.~\ref{subfig:nd-t2}. Whenever the
\rosnode{Behavior Planner} decides that it is time to brake and start backing up
into a parking space, it notifies the \rosnode{MPC Controller} with a
\rosmsg{Trajectory} that tells it to stop the vehicle. Simultaneously,
the \rosnode{Behavior Planner} sends the \rosnode{LGSVL Interface} a ``reverse
gear'' \rosmsg{State Command}. It would certainly be dangerous if the vehicle
interface were to attempt to change gear before reaching a full stop. To prevent
a serious error like this, the \awauto designers use a complicated state machine
in the implementation of the \rosnode{LGSVL Interface}
that ensures that all \rosmsg{Vehicle Command} messages are handled before
performing the gear change requested in the \rosmsg{State Command}. In our
opinion, this needlessly clutters the application logic with error-prone code. It would be much better
if the underlying framework would automatically provide proper alignment. 
\section{Design of \sysname}\label{sec:design}
The issues raised in Sec.~\ref{sec:motivation} can be automatically avoided if
the semantics of \lfshort are honored in a multi-process or distributed setting. However, there
are several non-trivial challenges involved in making a transition from
the currently implemented multi-threaded \lfshort runtimes to multi-process/distributed ones. In this section, we
discuss the design of our main innovation, the \sysname federated runtime for
\lfshort. We establish a transparent workflow that takes ordinary \lfshort
programs and turns them into \emph{federated} ones, in which reactors are mapped
to processes that can be deployed to remote machines, jointly constituting a \emph{federation}.
\subsection{Startup and Shutdown}\label{subsec:startup}
\subsubsection*{Startup} 
In \lfshort, there exists a \textbf{start tag} $g_s$ such that all logical
timelines for all reactors of a program start at $g_s$. In non-federated
\lfshort implementations, $g_s$ is set to a reading of a shared physical clock.
This approach does not extend well to a distributed system without access to a
shared clock, where clock synchronization is imperfect and communication delays have
to be accounted for. Instead, we implement a distributed consensus in which,
during a startup stage, a central coordinator called the RunTime Infrastructure
(RTI, the name adopted from HLA~\cite{dahmann1997department}) is started. Each
federate first synchronizes its physical clock with the RTI (see
Sec.~\ref{subsec:clock-sync} for details). Next, each federate reports a reading
from its physical clock and shares it with the RTI. The RTI collects a list
$\Phi_g$ of proposed start times and chooses $\max(\Phi_g) + d$ as the start
time for all federates, where $d$ is a value chosen automatically based on
measured communication delay and federation
size.
\subsubsection*{Shutdown} A deterministic shutdown is especially important
in safety-critical applications where external factors might require a timely,
but deterministic shutdown of the system (think of an advanced autopilot system
being turned off).
Since a federation is already running, the shutdown problem can be formulated as
a logical consensus. If a federate decides to stop the execution, it ceases
to advance its tag and notifies the RTI. The RTI will then ask other federates
for an appropriate final tag (which in turn causes the other federates to pause
processing events). After receiving all tags, the RTI picks the maximum tag and
sends it to all federates, which will then process all remaining events up to
and including this tag.

\subsection{Advancing Logical Time}\label{sec:advancing-time} To honor the
semantics of reactors, each federate must see events in tag order and not start
processing events with a tag larger than events that could later be produced by
upstream federates. In ordinary \lfshort programs, all reactors must finish
processing events with tag $g$ before any reactor may start processing an event
with tag $g' > g$. Such a barrier synchronization would be very costly in a
federated implementation. We instead have realized two more loosely coupled
coordination methods: our centralized and decentralized coordinators.



\subsubsection{Centralized Coordination}
For each federate $f$, the RTI keeps track of the following information:
\begin{compactEnumerate}
\item \LTC[f]: \textbf{Logical Tag Complete}. A record of the most recently
received tag from federate $f$ notifying the RTI that it has completed all
computation and sent all outgoing messages with that tag or less.  The value of
\LTC[f] is initially \NEVER, a special tag smaller than all other tags.
\item \NET[f]:  \textbf{Next Event Tag}. This is the most recently received tag
from a federate $f$ reporting the earliest event in its event queue. 
If a federate's event queue is empty, it will send \FOREVER, a special maximal
tag. If the RTI has not received a \NET message from the federate, the value is
\NEVER.
\item \TAG[f]:  \textbf{Tag Advance Grant}. This is the tag most recently sent
to federate $f$ to permit $f$ to advance its current tag to \TAG[f]. The value
of \TAG[f] is initially \NEVER.
\end{compactEnumerate}
Each federate $f$ conveys the first two quantities to the RTI in a \textbf{Next
Message Request} (\NMR[f]) message at the start of execution and at the
completion of each tag. The payload of the message is $(\LTC[f], \NET[f])$,
where $\LTC[f] < \NET[f]$. The very first \NMR message will carry \LTC[f] =
\NEVER because no event has been processed. The \TAG[f] is sent by the RTI to
each federate $f$ to permit it to advance its tag. At all times, the RTI and the
federates may have only partial information, and messages conveying these
quantities may be in flight and not have been recorded.

When the RTI receives an \NMR message from federate $f$, it may respond with a
\textbf{Tag Advance Grant} (\TAG[f]) message, possibly not immediately.
It may also send \TAG messages to downstream federates. Let $D(f)$ be the set of
immediate downstream federates (those that receive messages directly from $f$),
and let $U(f)$ be the set of immediate upstream federates (those that send
messages directly to $f$). 
Suppose the RTI receives an \NMR message from $f$ with payload $(\LTC[f],
\NET[f])$. First, it updates its data structure to register these two new
values. Then it does two things:
\begin{compactEnumerate}
\item For all $d \in D(f)$, let
\[
g_d = \min_{u \in U(d)} (\LTC[u] + a_{ud}),
\]
where $a_{ud}$ is the minimum ``after'' annotation on connections from federate
$u$ to $d$.
If $g_d > \TAG[d]$, then the RTI sets $\TAG[d] = g_d$ and sends a TAG message to
federate $d$ with payload $g_d$. This tells the downstream federate $d$ that it
may advance its current tag to $g_d$ and process any events or inputs that it
has received with that tag or any earlier tag. The RTI is sure to have forwarded
all messages to $d$ with tags equal to or less than $g_d$ because all federates
upstream of $d$, including $f$, have reported completion of execution at logical tag $g_d$ or
larger.
\item
The RTI determines whether it can grant a tag advance to $f$, in which case it
will send it a \TAG message. If $\TAG[f] \ge \NET[f]$, then there is nothing to
do because the RTI has already granted such a tag advance to $f$. Otherwise, if
$\TAG[f] < \NET[f]$, then the RTI needs to compute the \textbf{Earliest Incoming
Message Tag} (\EIMT[f]) for federate $f$. This is defined as follows:
\[
\EIMT[f] = \min_{u \in U(f)}(\min( \EIMT[u], \NET[u]) + a_{uf}) .
\]
If $U(f) = \emptyset$, then we define $\EIMT[f] = $ \FOREVER. Note that because
\EIMT appears on both sides, this is a system of equations. As long as the
logical delays satisfy $a_{uf} \ge 0$, it is easy to prove that there is a
maximal solution (which may have \FOREVER for some federates), even in the
presence of cycles, and we have implemented a simple iterative procedure for
finding that maximal solution.
If $\EIMT[f] \ge \NET[f]$, then let $\TAG[f] = \NET[f]$ and send a \TAG message
to $f$ with payload $\NET[f]$. Otherwise, the RTI does not reply. The federate
will have to wait until it gets a \TAG as a consequence of (1).
\end{compactEnumerate}


When a federate $f$ sends to the RTI an \NMR with payload $(\LTC[f], \NET[f])$,
it is stating that:
\begin{compactItemize}
\item
It will never again send a message with tag $g \le \LTC[f]$, and it would be an
error for it to receive any incoming message with such a tag (such a message is
said to be \textbf{tardy}, and a main task of the centralized controller is to
guarantee that no tardy messages occur).
\item
Until the federate receives a $\TAG[f] > \LTC[f]$, it will not send an outgoing
message with tag $g < \NET[f]$. 
Once it receives a $\TAG[f] > \LTC[f]$, then it can advance its current tag to
$\TAG[f]$ and send a message with tag $g \ge \TAG[f]$ (it can be $\TAG[f] +
a_{fd}$, where $a_{fd}$ is the ``after'' delay on the connection to $d$).
\end{compactItemize}

If federate $f$ has no physical actions, then \NET[f] is simply the tag of the
earliest event on its event queue, or \FOREVER if the event queue is empty.
However, 
if $f$ has physical actions, it can make no such promise until $T_f > \NET[f]$,
where $T_f$ is the current physical time at federate $f$, since an event might
appear with timestamp $T_f$.  
In this case, in order to permit downstream federates to advance time, federate
$f$ needs to repeatedly send \NMR messages as its physical time advances.  Here
there is an inherent tradeoff between network bandwidth and time granularity.
These messages need to be frequent, but not too frequent.

Fig.~\ref{fig:centralized_timeline} shows the sequence of messages under
centralized coordination that ensures alignment for the sub-architecture of
Autoware in Fig.~\ref{subfig:nd-t1}. The \code{MPC Controller} will not
process the received vehicle kinematic state message $\mathcal{V}_{g}$ until it
receives a \code{TAG($g$)} message from the RTI. The RTI in turn ensures that
the \code{MPC Controller} receives the \code{TAG($g$)} message only after it
has received the trajectory message $\mathcal{T}_{g}$, allowing the \code{MPC
Controller} to process $\mathcal{V}_{g}$ and $\mathcal{T}_{g}$ logically
simultaneously at tag $g$.





\begin{figure}[tb]
  \begin{center}
    \begin{tikzpicture}[font=\sffamily,scale=0.79]
    \footnotesize

    \draw[line width=1pt,-Latex] (0,20mm) -- (80mm,20mm) node[anchor=west, text width=17mm] {\scriptsize \lgsvl};
    \draw[line width=1pt,-Latex] (0,10mm) -- (80mm,10mm) node[anchor=west, text width=17mm] {\scriptsize \controller};
    \draw[line width=1pt,-Latex] (0,0mm) -- (80mm,0mm) node[anchor=west, text width=17mm] {\scriptsize \planner};
    \draw[line width=1pt,-Latex] (0,-10mm) -- (80mm,-10mm) node[anchor=west] {\scriptsize \rti};

    \node[pevent] at (3.13mm,20mm) (net_lgsvl) {};
    \node[pevent] at (12.5mm,20mm) (tag_lgsvl) {};
    \node[pevent, color=black!30!blue] at (15.5mm,20mm) (vse_lgsvl_planner) {};
    \node[pevent, color=black!30!blue] at (18.0mm,20mm) (vse_lgsvl_controller) {};
    \node[pevent] at (33.6mm,20mm) (ltc_lgsvl) {};

    \node[pevent, color=black!30!blue] at (18.00mm,0mm) (behavior_vse_from_lgsvl) {};
    \node[pevent] at (24.2mm,0mm) (net_planner) {};
    \node[pevent] at (47.1mm,0mm) (tag_planner) {};
    \node[pevent, color=black!30!blue] at (50.2mm,0mm) (trajectory_to_controller) {};
    \node[pevent] at (62.5mm,0mm) (ltc_planner) {};

    \node[pevent, color=black!30!blue] at (23.0mm,10mm) (controller_vse_from_lgsvl) {};
    \node[pevent] at (27.3mm,10mm) (net_controller) {};
    \node[pevent, color=black!30!blue] at (54.4mm,10mm) (trajectory_from_planner) {};
    \node[pevent, label={[text=black!60!green, xshift=-25mm, text width=50mm]45:\scriptsize Safe to process $\mathcal{V}_{g}$ and $\mathcal{T}_{g}$}, color=black!60!green] at (72.0mm,10mm) (tag_controller) {};

    \node[pevent] at (6.26mm,-10mm) (net_from_lgsvl_rti) {};
    \node[pevent] at (9.39mm,-10mm) (tag_to_lgsvl_rti) {};
    \node[pevent] at (27.3mm,-10mm) (net_from_planner_rti) {};
    \node[pevent] at (33.6mm,-10mm) (net_from_controller_rti) {};
    \node[pevent] at (39.8mm,-10mm) (ltc_from_lgsvl_rti) {};
    \node[pevent] at (43.0mm,-10mm) (tag_to_planner_rti) {};
    \node[pevent] at (65.6mm,-10mm) (ltc_from_planner_rti) {};
    \node[pevent] at (68.9mm,-10mm) (tag_to_controller_rti) {};

    \path[shortarrow] (net_lgsvl) edge [bend right=10, at start] node[anchor=south,inner sep=1mm, yshift=1mm] {\scriptsize {\tt NET($g$)}} (net_from_lgsvl_rti);
    
    \path[shortarrow] (tag_to_lgsvl_rti) edge [bend right=10] node[anchor=north,inner sep=1mm, near start, yshift=-8mm] {\scriptsize {\tt TAG($g$)}} (tag_lgsvl);
    
    \path[shortarrow] (vse_lgsvl_planner) edge [bend left=10, color=black!30!blue] node[anchor=south east,inner sep=2mm, text width=15mm] {} (behavior_vse_from_lgsvl);

    \path[shortarrow] (vse_lgsvl_controller) edge [bend left=10, color=black!30!blue] node[anchor=south,inner sep=1mm, yshift=5mm, xshift=-3mm, text width=17mm, align=center] {\scriptsize $\mathcal{V}_{g}$} (controller_vse_from_lgsvl);

    \path[shortarrow] (net_planner) edge [bend left=10] node[anchor=south,inner sep=0, yshift=6mm, xshift=-1mm] {\scriptsize {\tt NET($g$)}} (net_from_planner_rti);

    \path[shortarrow] (net_controller) edge [bend left=10] node[anchor=south,inner sep=1mm, yshift=9mm, xshift=-1mm] {\scriptsize {\tt NET($g$)}} (net_from_controller_rti);

    \path[shortarrow] (ltc_lgsvl) edge [bend left=10] node[anchor=south, inner sep=0, yshift=15mm, xshift=-2mm] {\scriptsize {\tt LTC($g$)}} (ltc_from_lgsvl_rti);
                                            
    \path[shortarrow] (tag_to_planner_rti) edge [bend left=10] node[anchor=north, inner sep=0, very near start, yshift=-3mm, xshift=2mm] {\scriptsize {\tt TAG($g$)}} (tag_planner);
    
    \path[shortarrow] (ltc_planner) edge [bend left=10] node[anchor=south,inner sep=0, near start, yshift=4mm] {\scriptsize {\tt LTC($g$)}} (ltc_from_planner_rti);

    \path[shortarrow] (trajectory_to_controller) edge [bend left=10, color=black!30!blue] node[anchor=north,inner sep=1mm, yshift=-6mm, xshift=1mm] {\scriptsize $\mathcal{T}_{g}$} (trajectory_from_planner);

    \path[shortarrow] (tag_to_controller_rti) edge [bend right=10] node[anchor=north,inner sep=0, yshift=-10mm] {\scriptsize {\tt TAG($g$)}} (tag_controller);  
    \end{tikzpicture}
  \end{center}
  \caption{Message sequence under centralized coordination that ensures alignment of vehicle kinematic state
  $\mathcal{V}_{g}$ and trajectory $\mathcal{T}_{g}$ for the architecture of 
  Fig.~\ref{subfig:nd-t1}. NMR messages are split into LTC and NET messages for the sake of clarity.}
  \label{fig:centralized_timeline}
  \vspace{-4mm}
\end{figure}
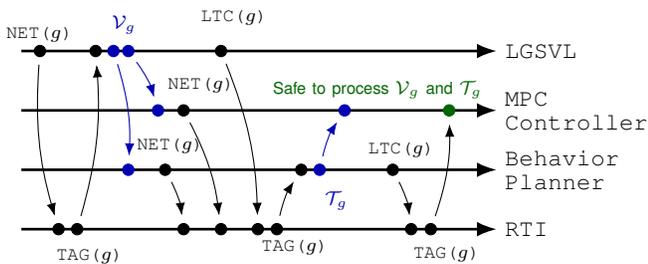

\subsubsection{Decentralized Coordination}

We give here a realization of PTIDES~\cite{Zhao:07:PTIDES, Zou:09:Ptides} with
substantial extensions.

For each federate $f_i$, we will derive $S_i \in \ttt$, a
\textbf{safe-to-process} (\textbf{STP}) offset. Given $S_i$, a federate will not
be able to advance to any tag $g = (t, m)$ until its physical time $T_i$
satisfies
\begin{equation}\label{eq:stp}
T_i \ge t + S_i .
\end{equation}
where $T_i $ is a measurement of physical time on the platform executing $f_i$.
If $f_i$ includes any physical action, then $S_i \ge 0$. Otherwise, it can be
positive, negative, or zero. The purpose of $S_i$ is to ensure that by the time
the physical clock satisfies (\ref{eq:stp}), all events that may have originated
on some other federate with tags less than $g$ have previously arrived at $f_i$.
This ensures that events can be processed in tag order.

Next, we discuss the assumptions needed to calculate $S_i$.
\paragraph{Launch Deadline} Assume that a reaction $r$ invoked at tag $g$ in
$f_i$ is able to send a message to another federate $f_j$. It does this simply
by writing to an output port of $f_i$. We assume that $r$ is invoked before
physical time $T_i$ exceeds $t + D_{ij}$, where $D_{ij}$ is the launch
deadline. As mentioned in Sec.~\ref{subsec:lf-background}, violation of such a
deadline can be detected using existing semantics of \lfshort. It is worth
noting that our runtime interprets a lower deadline as having a higher priority.
Also note that it must be true that $D_{ij} \ge S_i$ because no reaction in
\lfshort can be invoked at tag $g = (t,m)$ before physical time $T_i \ge t +
S_i$. 

\paragraph{Launch Lag} We define the launch lag as
\begin{equation}\label{eq:launchlag}
L_{ij} = D_{ij} - S_i .
\end{equation}
The launch lag $L_{ij}$ represents the physical time that elapses between the
start of execution of the step at tag $g$ and the invocation of reaction $r$.
It must include the execution time of any reaction(s) that are invoked before
$r$ at this tag plus any scheduling overhead, and hence it must be nonnegative.
The launch lag may be zero if this overhead is negligible.



\paragraph{Communication latency bound}
We also need to assume a bound $N_{ij}$ on communication (either
inter-process or network) latency. This is defined as
the maximum physical time (by the clock at federate $f_i$) that passes between
the invocation of the reaction $r$ that sends the message and the receipt of the
message at federate $f_j$. This time includes not just the propagation time
through the communication channel (sockets in our case), but also the execution time of $r$ and any overhead in the
network stack.

\paragraph{Clock synchronization error bound}
Finally, we also need to assume a bound $E_{ij}$ on the clock synchronization
error between federates $f_i$ and $f_j$. That is, the physical clocks of $f_i$
and $f_j$ do not drift apart by more than $E_{ij}$.


\noindent\textbf{Calculating the Safe To Process Offset.} Assume that $f_i \in F$ sends an event to federate $f_j \in F$ with tag $g =
(t,m)$. 
With the aforementioned bounds, $f_j$ will receive the message before $f_i$'s
physical clock reaches $t + D_{ij} + N_{ij}$ and before its own physical clock
reaches $t+D_{ij} + N_{ij} + E_{ij}$.

The connection from $f_i$ to $f_j$ may alter the tag (using the \code{after}
keyword), incrementing it by $a_{ij} \ge 0$. Since the event is launched by a
reaction invoked at $t$, the event with timestamp $t' = t+a_{ij}$ will be
received by $f_j$ before its physical clock exceeds $t+D_{ij}+ N_{ij} +
E_{ij}$. Equivalently, $f_j$ receives an event with tag $g' = (t', m')$ before
its physical clock exceeds
\begin{equation}\label{eq:messagedelay}
T_j = t' + D_{ij} + N_{ij} + E_{ij} - a_{ij}.
\end{equation}
If $a_{ij}$ is large enough, it can even be true that $T_j < t'$, in which
case, at physical time $T_j$,
the receiving federate will have received all messages with timestamps $t \le
T_j$.

We can now generalize to any number of federates. First, let $\alpha_{ij}$ be
the \textbf{minimum delay} (specified using \code{after}) over all connections
from $f_i$ to $f_j$. Let
\begin{equation}\label{eq:matrix}
M_{ji} = \max (0, D_{ij} + N_{ij} + E_{ij} - \alpha_{ij}).
\end{equation}
Here, if $i = j$, $E_{ij} = 0$, and $N_{ij}$ is a bound on the latency with
which the federate sends messages to itself (this could be zero as well). If
there is no connection from $f_i$ to $f_j$, then $\alpha_{ij} = \infty$.
$M_{ji}$ represents a \emph{relative} safe-to-process offset for $f_j$ with respect to $f_i$.

We can now write down an expression for the safe-to-process offset $S_j$,
\begin{equation}\label{eq:stfp1}
S_j = \max_{i \in F}(M_{ji}).
\end{equation}
However, from (\ref{eq:launchlag}), we have
\[
D_{ij} = S_i + L_{ij},
\]
so the safe to process offsets appear on both sides of this equation. Therefore,
we have a system of equations that must be solved. Rewriting (\ref{eq:matrix})
to make this explicit, we get,
\[
M_{ji} = \max (0, S_i + L_{ij} + N_{ij} + E_{ij} - \alpha_{ij}).
\]
To simplify this, let
\[
X_{ij} = 
L_{ij} + N_{ij} + E_{ij} - \alpha_{ij}
\]
and write
\[
M_{ji} = \max (0, S_i + X_{ij}).
\]
We can now write
\begin{equation}\label{eq:tosolve}
S_j = \max_{i \in F}(M_{ji})
= \max(0, \max_{i \in F}( S_i + X_{ji}) ).
\end{equation}
We now have $n$ equations in $n$ unknowns.


\noindent\textbf{Max-Plus Formulation.} Equation (\ref{eq:tosolve}) can be written
compactly and solved easily using a Max-Plus formulation
\cite{Baccelli:92:MaxPlus}. Let $\mathbf{X}$ be a matrix where the $i,j$-th
element (row $i$, column $j$) is $X_{ij}$. Let $\mathbf{S}$ be a column vector
where the $i$-th element is $S_i$. Then note that in Max-Plus algebra,
\[
\mathbf{J} = \mathbf{X}\mathbf{S}
\]
is a column vector where the $j$-th element is $\max_{i \in F}( S_i + X_{ji})$.
Let $\mathbf{O}$ be a column vector where the $j$-th element is $0$. Then note
that (\ref{eq:tosolve}) can be rewritten in Max-Plus algebra as
\begin{equation}\label{eq:maxplus}
\mathbf{S} = \mathbf{O} \oplus \mathbf{J} = \mathbf{O} \oplus \mathbf{X}\mathbf{S}.
\end{equation}
From \cite{Baccelli:92:MaxPlus} (Theorem 3.17), if every cycle of the matrix
$\mathbf{X}$ has weight less than zero, the unique solution of this equation is
\[
\mathbf{S} = \mathbf{X}^* \mathbf{O},
\]
where the \textbf{Kleene star} is (Theorem 3.20)
\[
\mathbf{X}^* = \mathbf{I} \oplus \mathbf{X} \oplus \mathbf{X}^2 \oplus \cdots
\]
and that this reduces to
\[
\mathbf{X}^* = \mathbf{I} \oplus \mathbf{X} \oplus  \cdots \oplus \mathbf{X}^{n-1},
\]
where $n$ is the number of federates. If there are cycles with zero weight, but
all weights are nonpositive, then there is a solution and a unique minimal
solution.

The requirement that the cycle weights be nonpositive is present in the original
PTIDES. 
For a federate $j$, \sysname will
optimistically replace $X_{ij}$ with $T_j - t$ for a given tag $g = (t, m)$ if
the status of events for all connections from $i$ are known at $g$.

\subsection{Fault Handling}
Without exception, the correctness of a system implementation is predicated on
certain assumptions. For example, operating conditions, such as temperature or
humidity, need to be within a certain range. Typically, violation of the
assumptions leads to a fault condition. In \lfshort, with its sophisticated model
of time, special attention is given to faults related to broken assumptions
regarding time. Specifically, one can associate a \textbf{deadline} with a
particular reaction, along with a fault handler. The fault handler gets invoked
\emph{instead} of the reaction in case it is not ready to execute by the
specified deadline. When the centralized coordinator is used in \sysname, the
deadline fault handler can be used by the user to detect and react if the underlying
communication channel is partitioned or assumptions about execution times are
violated.

When the decentralized coordinator is used in \sysname, there is another type of time-related
fault that can occur. Specifically, a message from one federate to another
could be \textit{tardy}, meaning it arrives after the receiving federate has
already advanced to a tag greater than the incoming message. This can only
happen, of course, if the asserted STP offset was not large enough to account
for upstream delays, meaning that an \textbf{STP violation} occurred. To allow
the user to react to such a violation, we added a
new syntax to \lfshort for the specification of the STP, along with a handler to
be invoked when it is violated, emulating the style of the existing deadline
construct.
\section{Implementation}\label{sec:implementation}
\begin{figure}[t]
  \centering
  \includegraphics[width=.9\linewidth, trim=0cm 5.8cm 6.8cm 0, clip]{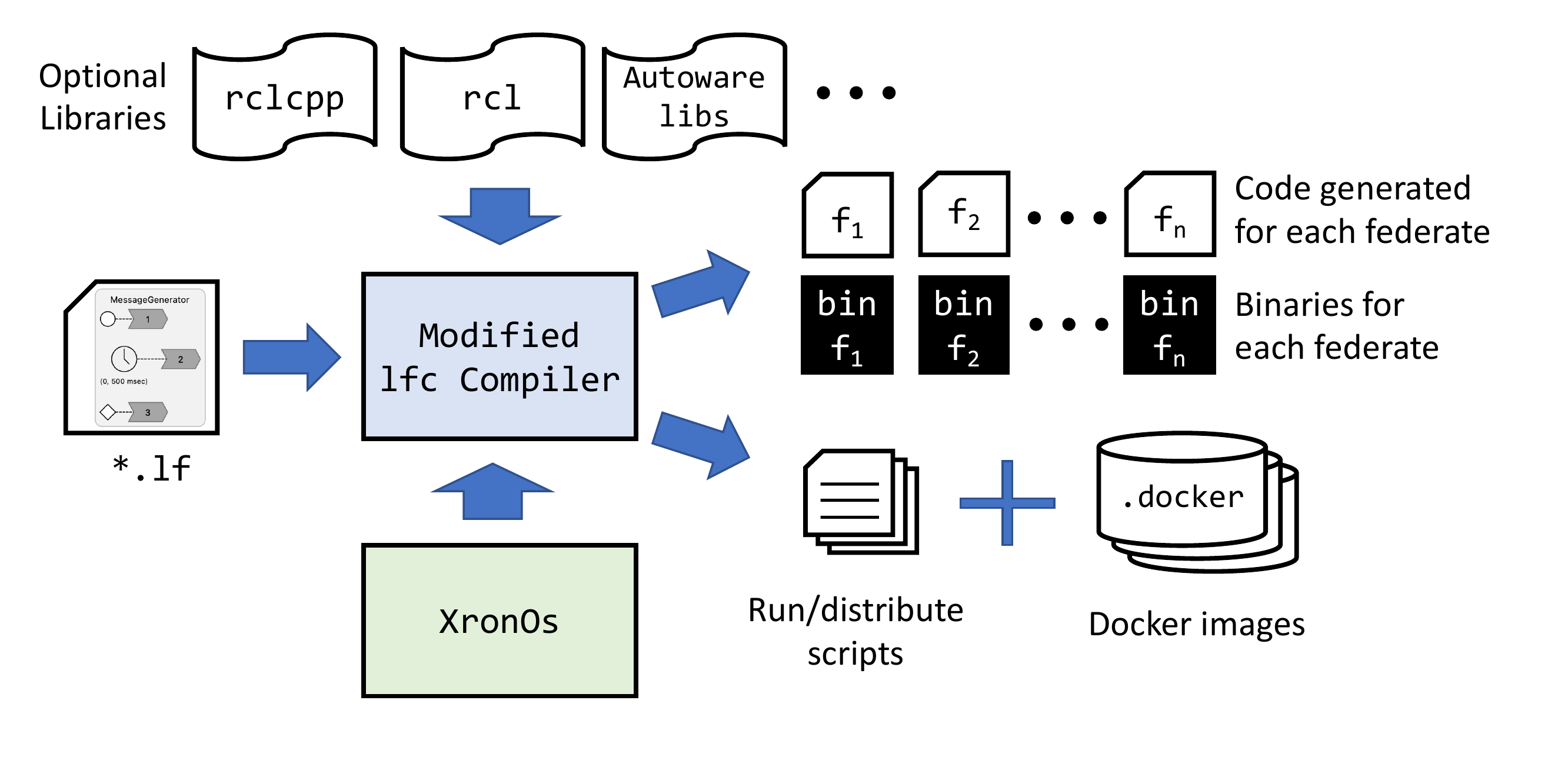}
  \caption{\sysname and \lfshort: compilation process.}
  \label{fig:xronos}
  \vspace{-3mm}
\end{figure}
\sysname is a runtime implementation for \lf, an open-source coordination language
built on the basis of the reactor model. 
At the time of writing, \lfshort supports C, C++, Python, TypeScript, and Rust as
targets.
Any \lfshort program using the C/C++ or Python targets can run on \sysname. The functionality
for this gets enabled simply by changing the \keyword{main} keyword to
\keyword{federated}. Support for \sysname in the TypeScript
target is under development. Fig.~\ref{fig:xronos} illustrates the
integration between \sysname and \lfshort as well as the steps involved in
constructing a federated program.

\subsection{Additions to the Language} The example \lfshort code in
Fig.~\ref{fig:example-federated} highlights our additions to the \lfshort
grammar. We added several \emph{target properties}, which are meant to configure
a program and its execution environment in various ways. For instance, we added
an option to select a coordination type, an option to create Docker images, and
several configuration options pertaining to the clock synchronization mechanism
provided by \sysname. We further add the \keyword{serializer} keyword to
the language to allow the programmer to specify a serialization mechanism for
data exchanged between federates. The \keyword{at} keyword was
introduced for mapping federates to specific hosts. Finally, we added a reserved
reactor parameter called \code{stp\_offset} that can be used to specify an STP
offset, as well as syntax for defining STP violation handlers.

\subsection{Federated Runtime}\label{subsec:fedruntime}
The federated runtime is written in C and uses sockets for communication. The
entire runtime is approximately 8100 lines of code. 
\sysname works in conjunction with our modified C and Python code generators that
automatically treat reactor instances in the top-level reactor as federates and
map them to independent processes. The code generators also transform
connections between federates such that the communication gets routed through a
socket connection using special sender and receiver reactions. 
Depending on the coordination type, communication will either happen directly
between federates 
(decentralized) or through the RTI (centralized). For logical connections, the
receiver reaction is triggered by a logical action that is scheduled by the
federated runtime upon message receipt. The tag of the event that triggers the
receiver reaction is determined by the tag that is sent along with the incoming
message.
For physical connections, the receiver reaction is triggered by a physical
action. The tag associated with the triggering event is thus based on physical
time as measured by the receiver. Hence, messages sent along physical
connections do not carry a tag. 
This type of connection is the closest equivalent to the connections used in ROS
and MQTT.

\begin{figure}[t]
  \centering
  \begin{lstlisting}[style=framed,language=LF,escapechar=|]
target C {
  coordination: decentralized,  // Or "centralized"
  clock-sync: on,     // Turn on runtime clock sync
  clock-sync-options: {
      period: 5 msec, // Clock sync period
      trials: 10,     // Num. of msgs used in clock sync
      attenuation: 10 // Stabilize clock sync
      ... other parameters
  },
  docker: true,       // Produce a docker image
  timeout: 10 secs,   // Distributed timeout
  tracing: true       // Distributed tracing
}
reactor Source {
  output out:int
  reaction(startup) -> out {=
      SET(out, 1);    // Send data over socket
  =}
}
reactor Destination (stp_offset: time(7 usec)) {
  input in:int;
  reaction(in) {=
      info_print("%d", out); // Print the received data
  =} STP {=
      // Handle input that violates tag order
  =}
}
federated reactor Example at user1@host1 {
  c = new Source()      at user2@host2
  d = new Destination() at user3@host3
  c.out -> d.in serializer "native" // Or ros2, proto
}
  \end{lstlisting}
  \vspace{-4mm}
  \caption{An example federated program.}
  \label{fig:example-federated}
  \vspace{-3mm}
\end{figure}

\subsection{Physical Clock Synchronization}\label{subsec:clock-sync}
\sysname
can work with clock synchronization based on NTP (default in most systems) or
the higher precision PTP~\cite{Eidson:06:1588,EidsonStanton:15:Time}.
We also provide a built-in clock synchronization implementation realized using
the technique of Geng et al.~\cite{GengEtAl:18:ClockSync}. This ensures that
federated execution initializes as expected even if the host system lacks proper
configuration or means of clock synchronization.
\sysname also optionally corrects for clock drift during execution.

\subsection{Reusing MQTT and ROS Libraries and Nodes}
\lfshort is already designed to seamlessly incorporate (legacy) target code and
reuse existing libraries. To ease the transition to \sysname, we have prepared a
number of examples\footnote{In a private repository to keep anonymity.} that
demonstrate how to integrate existing ROS or MQTT applications into
\sysname-based federations.
We also added \code{cmake} build support to the C target of our modified
\code{lfc} compiler and introduced target properties for customizing the
\code{cmake} build configuration to allow for more convenient integration
with the \code{colcon} build system used by ROS 2.




\subsection{Serialization}\label{subsec:serialization}
To send data from one federate to another, the data must be converted to a byte stream
(it must be serialized) and converted back to the appropriate data type at the receiver
(it must be deserialized).
\sysname currently supports three serialization schemes:
\keyword{native}, \keyword{proto}, and \keyword{ros2}.
In C, the \keyword{native} technique directly copies the memory map of the
data to the byte-array that is sent over the socket. This method of
serialization can be dangerous and only works if the data type is Plain
Old Data (POD)~\cite{Black:1998:Dictionary} and if the memory format and
endianness are the same for the sender and receiver. Nonetheless, this method is
fast, and thus useful for some embedded applications. In Python, the \code{pickle}\footnote{\href{https://docs.python.org/3/library/pickle.html}{https://docs.python.org/3/library/pickle.html}}
module is used to perform \keyword{native} serialization. The
\keyword{proto} serializer uses
Protobufs,\footnote{\href{https://developers.google.com/protocol-buffers/}{https://developers.google.com/protocol-buffers/}}
where \code{libproto} or its variants are automatically used to serialize and
deserialize data. As with any proto-based serialization, the data format is
expected to be stored in a \code{.proto} file which is then compiled and made
accessible in the program. Lastly, \keyword{ros2} uses the
serialization framework of
\code{rclcpp}.\footnote{\href{https://github.com/ros2/rclcpp}{https://github.com/ros2/rclcpp}}
Only ROS 2 messages, services, and actions are supported as a data format. This
functionality is particularly important for porting legacy code from ROS 2 to
\sysname, enabling developers to reuse ROS 2 message and service data types
(e.g., \code{PointCloud2}).

\subsection{Deployment}

We added support for a deployment strategy based on
Docker~\cite{merkel2014docker} by
generating Docker images for each federate, if requested, and a Docker compose
file that simplifies the task of starting the federation, even in a hybrid
heterogeneous system.






\section{Evaluation}\label{sec:evaluation}
We first discuss our port of \awauto and measure the error rates under \sysname
for the scenarios that were discussed in Sec.~\ref{sec:motivation}.
We also compare the end-to-end latency of \awauto in ROS 2 against our port to
\sysname. 
Subsequently, we supplement our performance evaluations with a set of microbenchmarks to
measure the impact of different subsystems in \sysname, such as the built-in Protobufs
serialization mechanism, on maximum throughput.

\begin{figure}[t]
    \centering
	\begin{minipage}[t]{0.47\linewidth}
        \includegraphics[width=\linewidth]{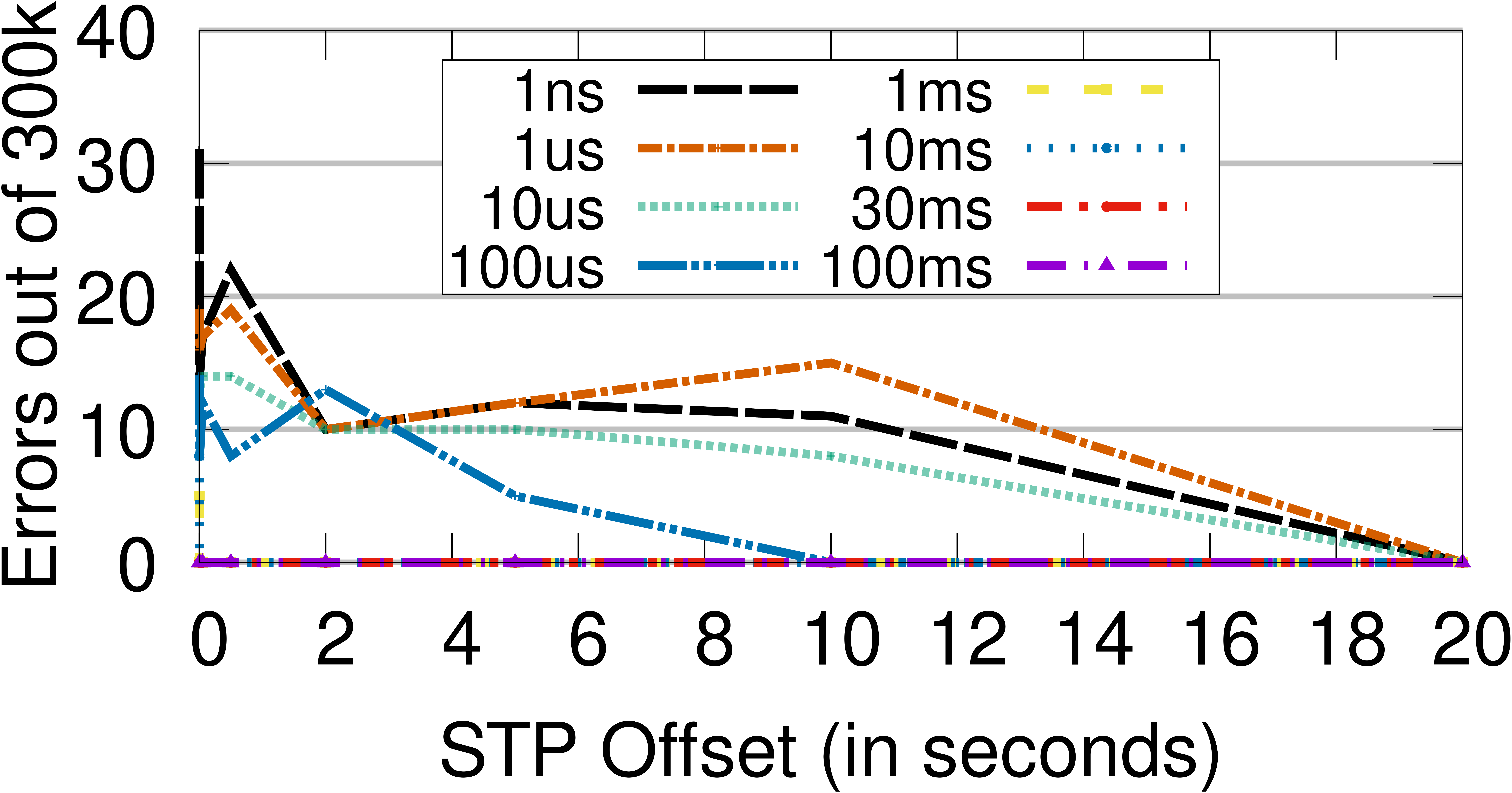}
        \captionof{figure}{Relationship between STP offset and error rate under
        decentralized coordination for the sub-architecture of Fig.~\ref{subfig:nd-n2n}.}
    \label{fig:autoware-n2n}
	\end{minipage}\hfill
	\begin{minipage}[t]{0.47\linewidth}
        \includegraphics[width=\linewidth]
        {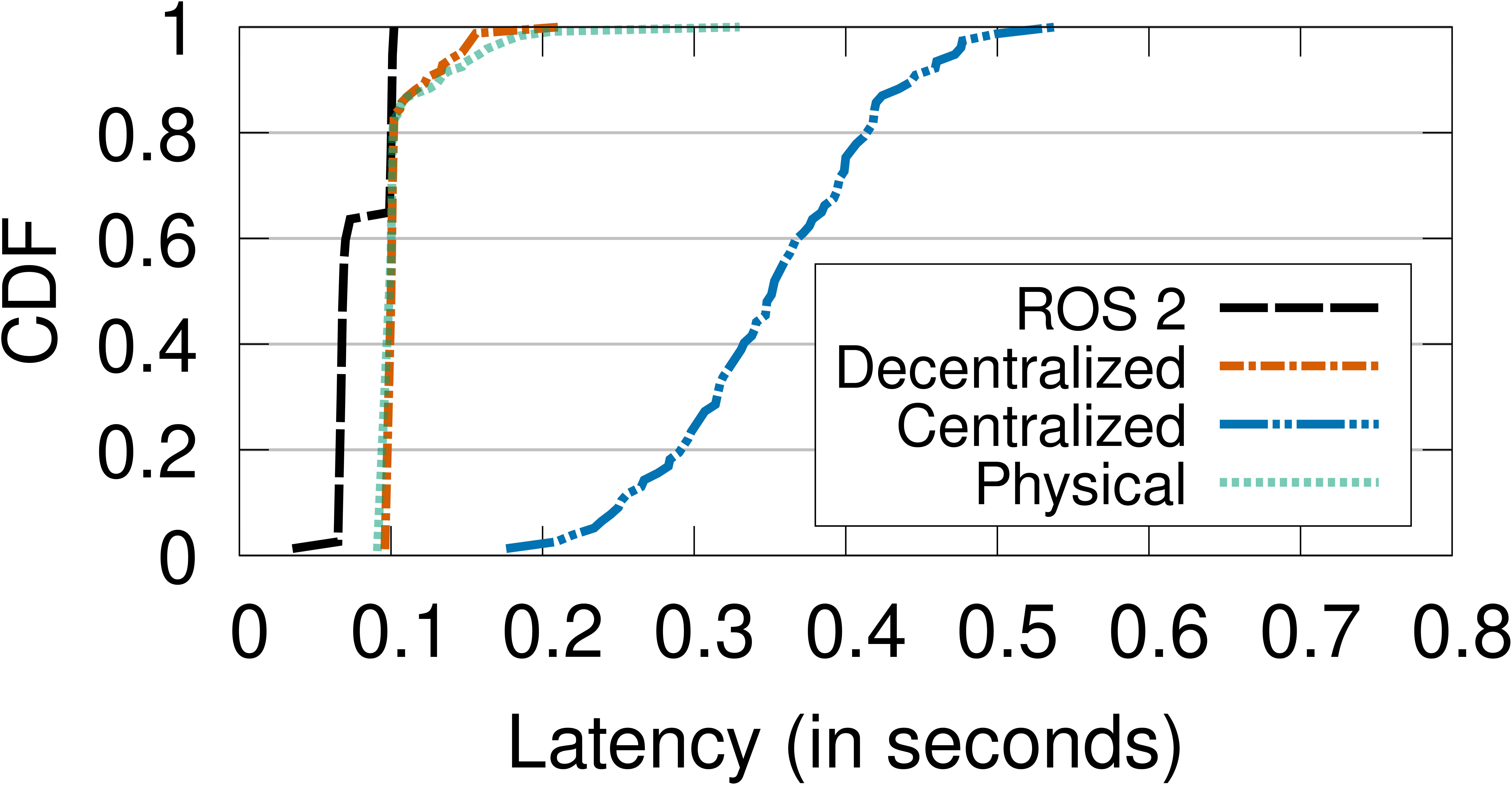}
        \captionof{figure}{Cumulative distribution function (CDF) 
        of end-to-end latencies for calculating a new lane trajectory in \awauto under 
        ROS 2 and \sysname.}
        \label{fig:autoware-e2e}
	\end{minipage}
    \vspace{-4mm}
\end{figure}

\subsection{\awauto}\label{sec:eval:autoware} We ported \awauto release
1.0 to \sysname, including its autonomous valet parking functionality. We
use our port to measure error rates under both realistic and unrealistic periods
for the sub-architecture of Fig.~\ref{subfig:nd-n2n}. First, we verified that centralized
coordination yields zero errors over 300k test runs. With decentralized coordination,
errors become possible, but we found no errors for periods down to one millisecond
(with an STP offset of 5 ms).
With periods below one millisecond, errors begin to appear, but, unlike the ROS 2
implementation, they are detectable (as STP violations). Moreover, by increasing the
STP offset, the error rate can be reduced. We found, however, that with periods of
100 $\mu$s and below, the STP offset had to become quite large (several seconds)
for the error rate to drop to zero, as shown in Fig.~\ref{fig:autoware-n2n}.
Such periods are unrealistic for this application. For applications with such
periods, more specialized hardware and networking may be required to eliminate
errors.


We would expect that such an improvement in reliability would incur a cost,
and, indeed, under centralized coordination, the cost is significant for this
application. Under decentralized coordination, we find that the cost in
end-to-end latency is considerably lower.

To evaluate the cost for this application, we measured the end-to-end latency
from sensor measurements (at the interface to the LGSVL simulator) through the
production of a new trajectory in the lane planner to the construction of a
vehicle command to feed back to the LGSVL simulator. First, we note that our
evaluation still involves two unported ROS 2 components, the LGSVL simulator
itself and the built-in tf2 library, which performs frame transformations.
These components are not part of \awauto, but they are needed to construct a
full simulation. Also, note that the end-to-end latency includes not just
communication overhead, but also the computations needed for planning. The goal
of the experiment is to compare the overall behavior of \sysname compared to ROS
2 for a realistic application.

In Fig.~\ref{fig:autoware-e2e}, we have plotted a cumulative distribution
function (CDF) as a function of latency.


Under centralized coordination, the latency is much higher, suggesting that
centralized coordination would not be a good choice for this application. This
is not surprising because, with centralized coordination, multiple control
messages need to be exchanged for transmission of each useful data message, and
all communication between federates goes through the RTI. The microbenchmarks,
which we consider next, also show that the communication overhead of centralized
coordination is quite high, suggesting that this strategy would only be
acceptable if occasional errors cannot be mitigated.

However, under decentralized coordination, the cost in end to end latency
becomes considerably more manageable, but not negligible. Nonetheless, we
conjecture that for safety-critical distributed embedded applications, the overall
reduction in the error rate could justify this cost.

To further clarify the source of the additional overhead in \sysname, we
measured the end-to-end latency of our port using all physical connections,
where, as previously explained in Sec.~\ref{subsec:lf-background}, the inherent
determinism of \lfshort is sacrificed. As depicted in Fig.~\ref{fig:autoware-e2e},
the resulting latency roughly matches that of the decentralized coordination.
This suggests that the added logic in decentralized coordination that ensures
determinism is not the main source of the added latency. We leave further
optimizations of the \sysname runtime as future work.


\subsection{Microbenchmarks}\label{subsec:microbenchmarks}
To supplement our results, we consider three distinct communication patterns, shown in
Fig.~\ref{fig:microbenchmarks}, inspired by patterns introduced by Lee and 
Lohstroh~\cite{LeeLohstroh:2021:TimeForAll}, to evaluate the throughput of
\sysname as a middleware.
We develop microbenchmarks for these patterns in \sysname, ROS 2, and MQTT and
measure the maximum throughput, comparing our decentralized coordinator against
the state-of-the-art to establish a baseline for comparison. We also measure the
throughput of \sysname under the centralized coordinator and
measure the impact of various features in \sysname on throughput. We
run our tests on three platforms, a desktop PC (the same system used to obtain
the results in Sec.~\ref{sec:motivation}), an NVIDIA Jetson AGX, and a hybrid
scenario where nodes/federates are distributed across the PC and the AGX.
\begin{figure}[t]
    \centering
    \begin{subfigure}[b]{.33\linewidth}
        \hspace{2em}
        \includegraphics[width=0.62\linewidth]{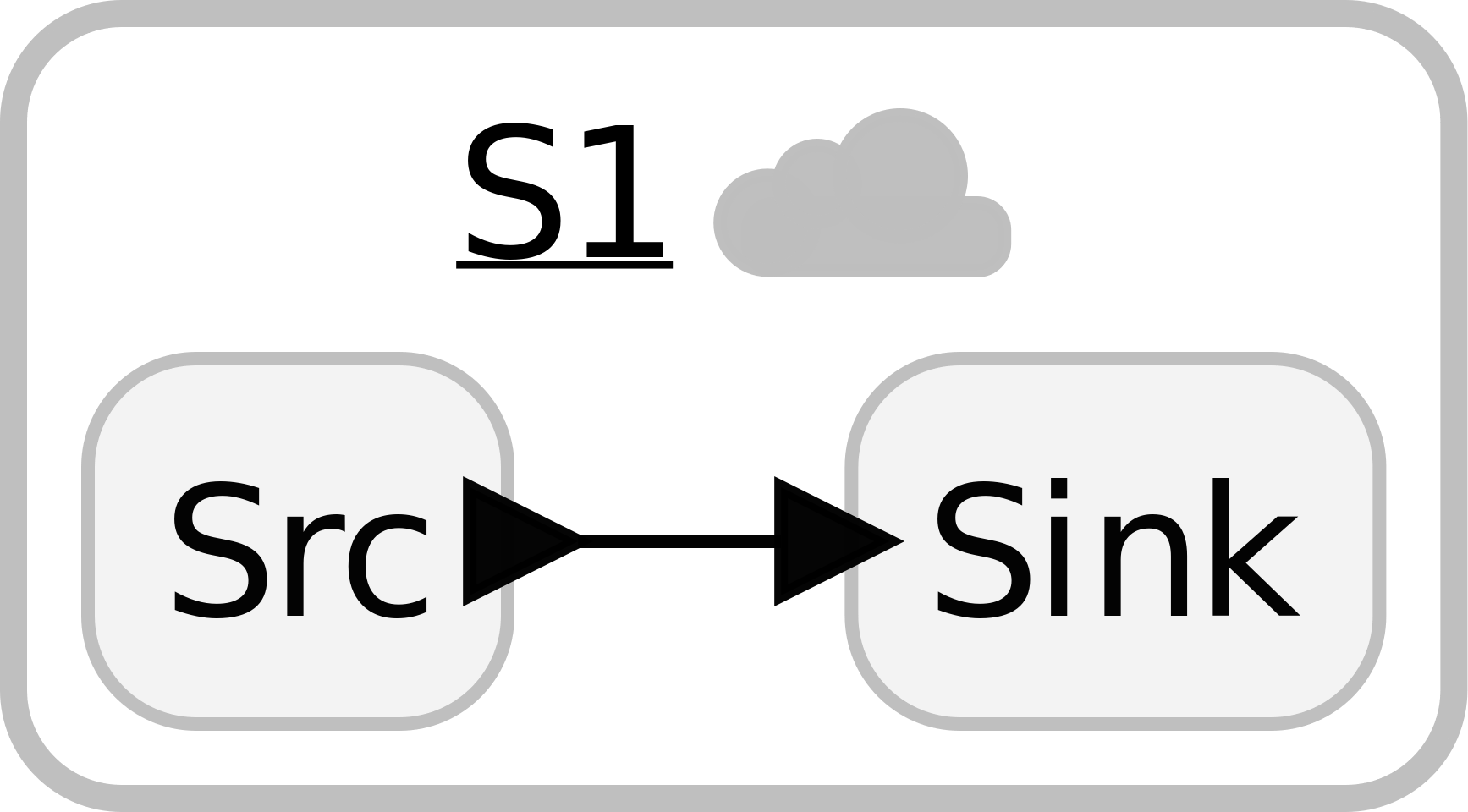}
    \end{subfigure}\hfill
    \begin{subfigure}[b]{0.33\linewidth}
        \hspace{1em}
        \includegraphics[width=.62\linewidth]{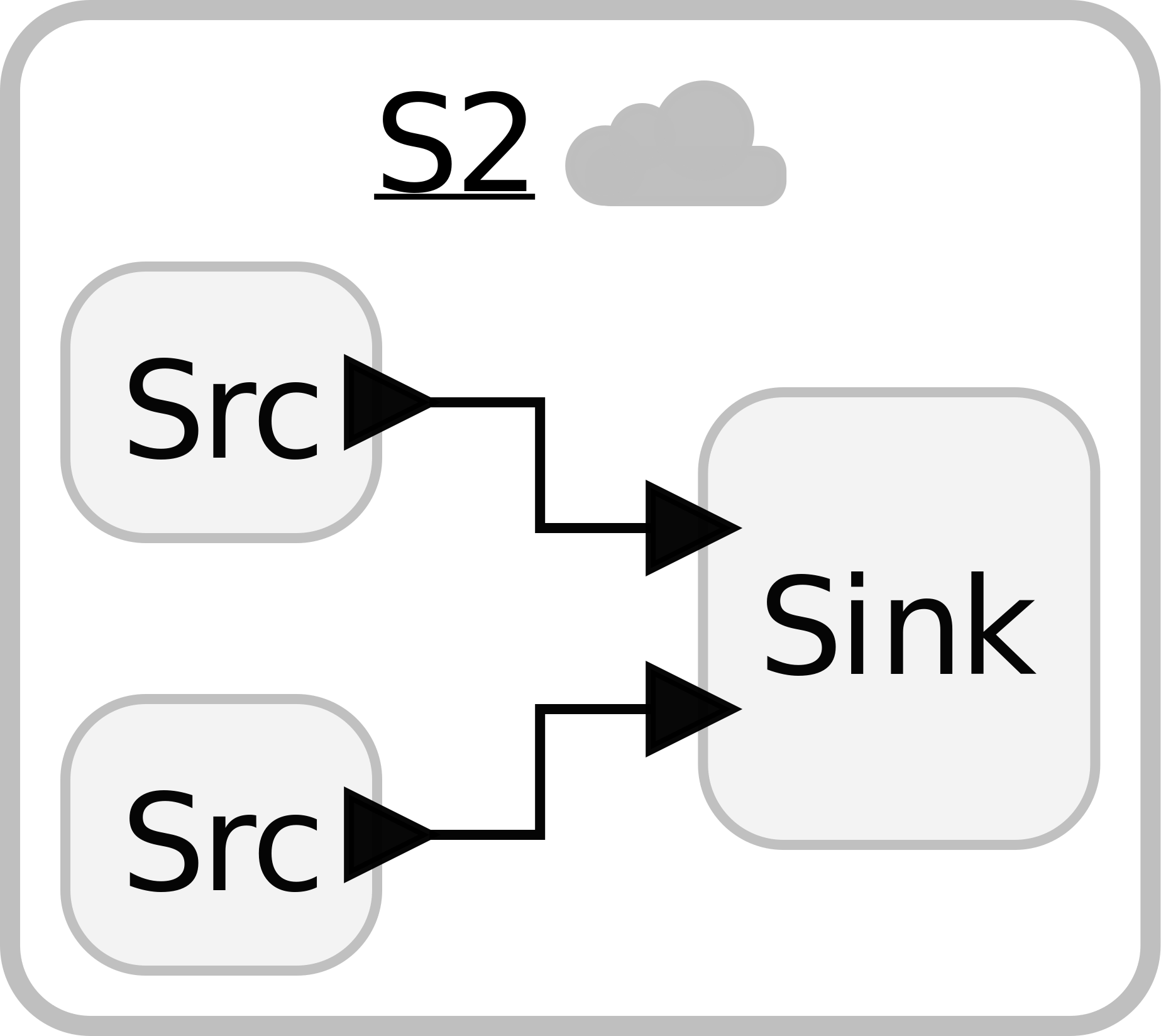}
    \end{subfigure}\hfill
    \begin{subfigure}[b]{0.33\linewidth}
        \includegraphics[width=.84\linewidth]{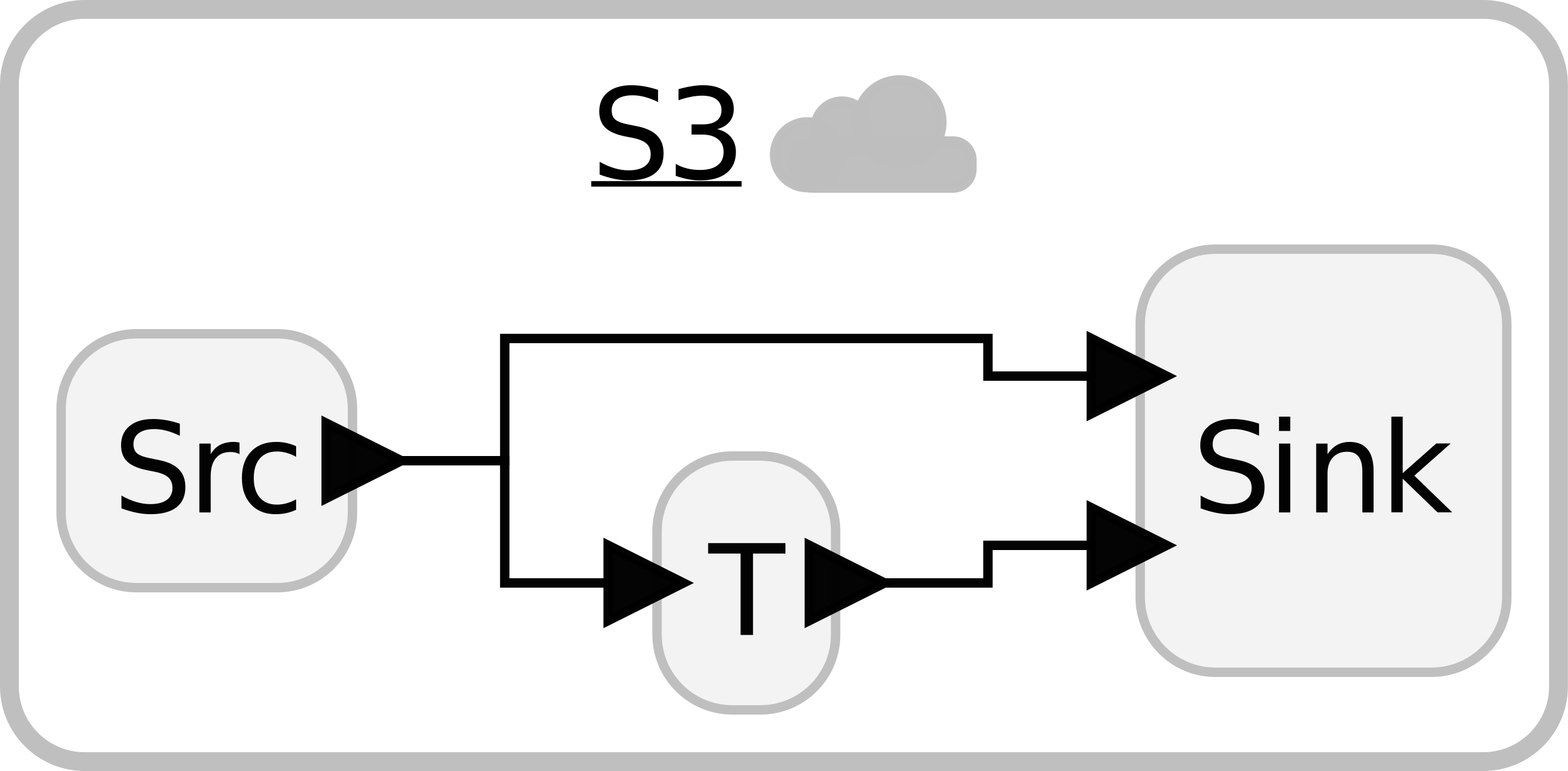}
    \end{subfigure}
    \caption{Microbenchmark communication patterns.}
    \label{fig:microbenchmarks}
    \vspace{-4mm}
\end{figure}

\begin{figure}[t]
    \centering
    \begin{minipage}[b]{\linewidth}
        \centering
        \includegraphics[width=0.7\linewidth]{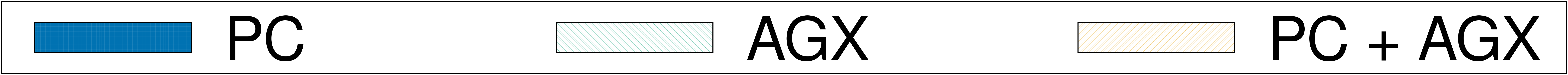}
    \end{minipage}\\
    \vspace{1mm}
    \begin{minipage}[b]{\linewidth}
      \centering
      \begin{subfigure}[b]{0.32\linewidth}
          \includegraphics[width=\linewidth]{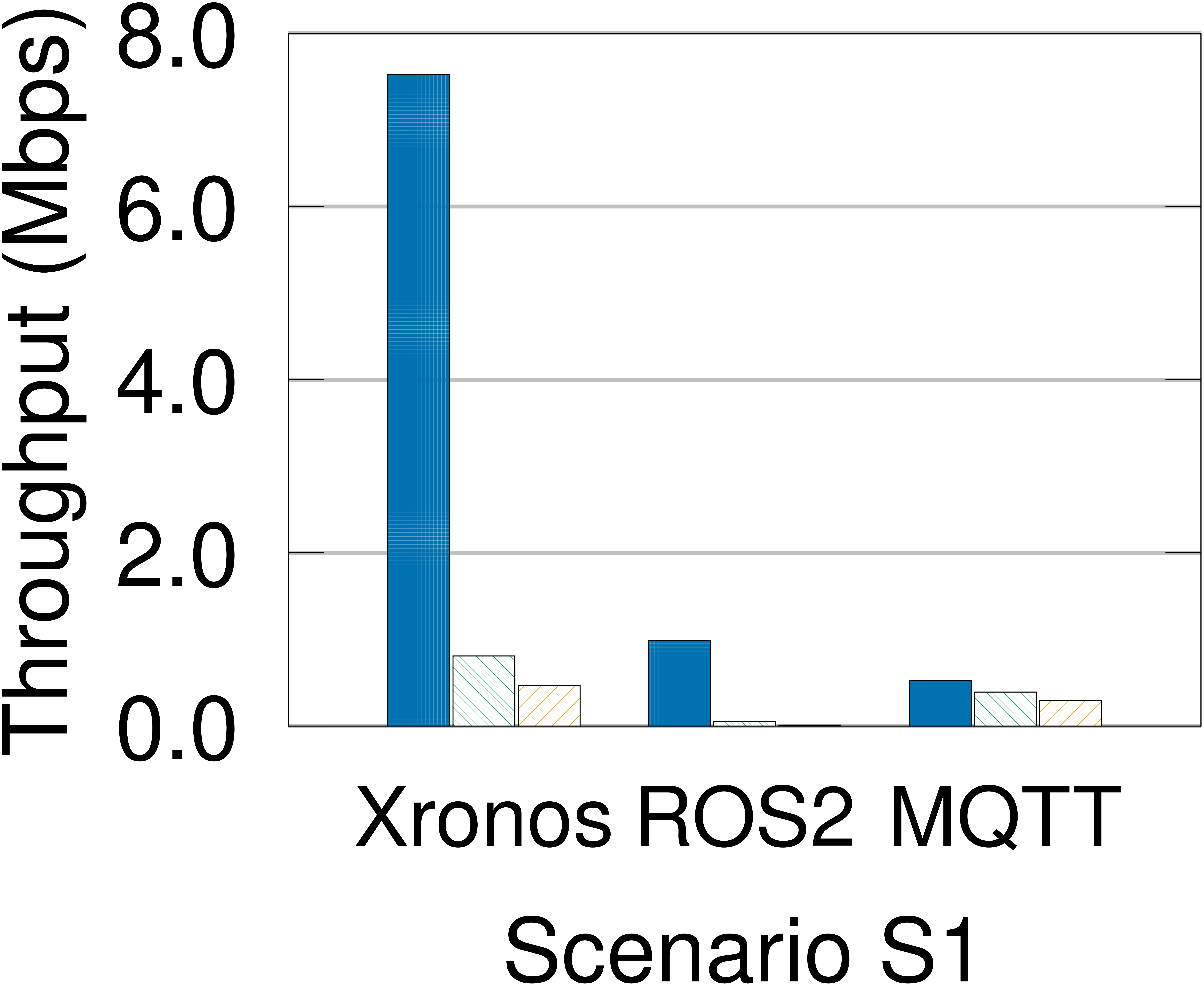}
      \end{subfigure}\hfill
      \begin{subfigure}[b]{0.32\linewidth}
          \includegraphics[width=\linewidth]{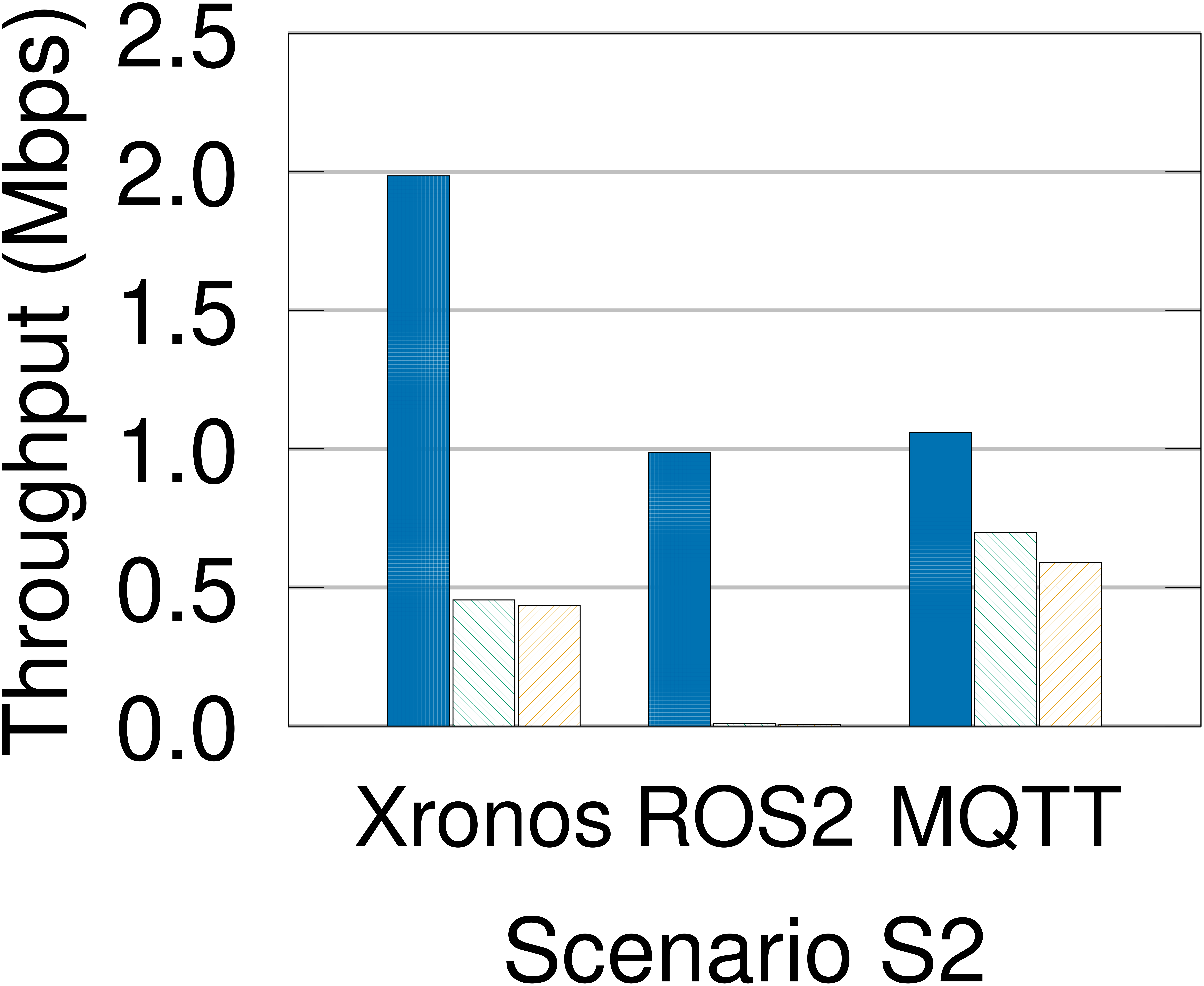}
      \end{subfigure}\hfill
      \begin{subfigure}[b]{0.32\linewidth}
          \includegraphics[width=\linewidth]{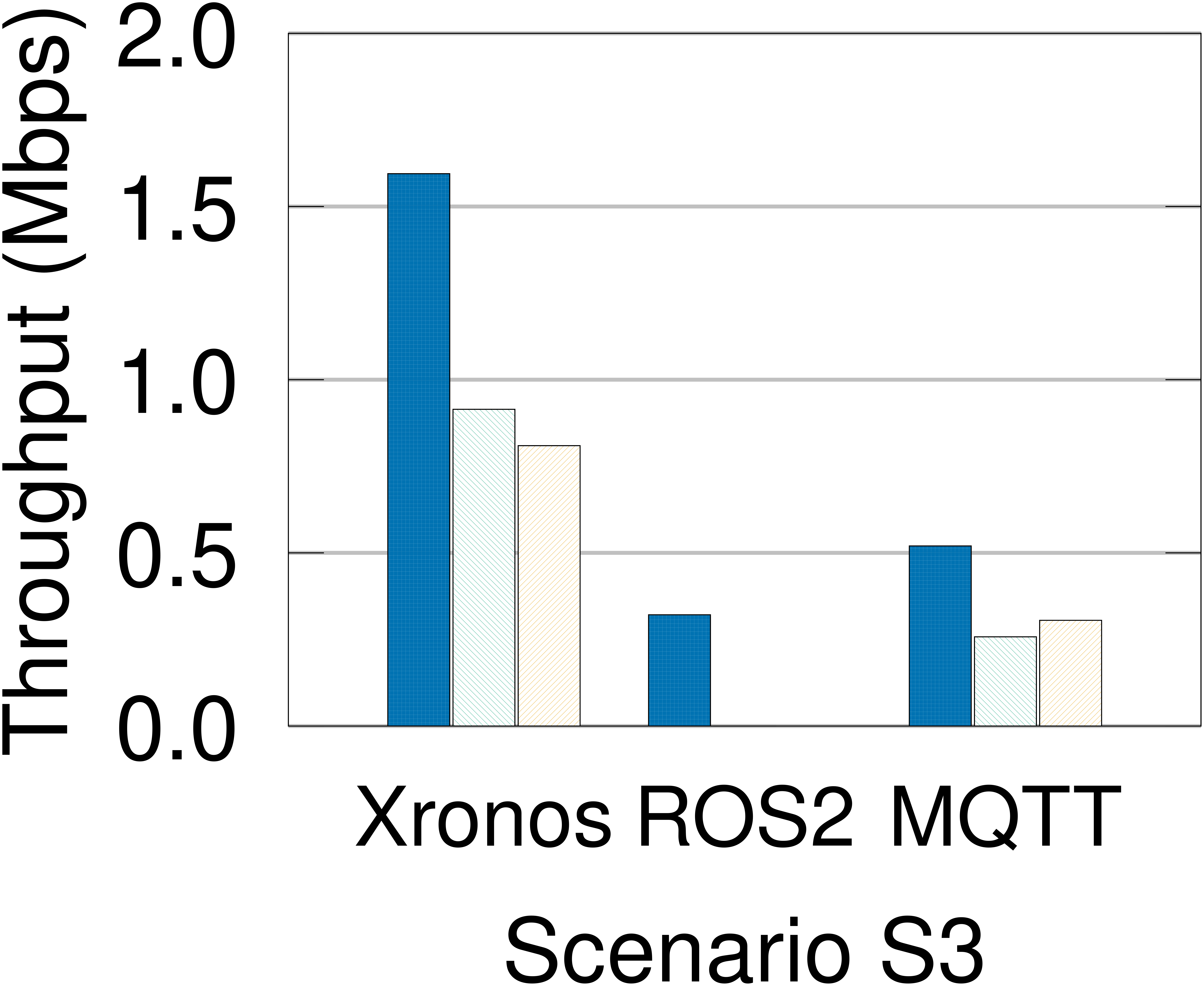}
      \end{subfigure}
      \caption{Comparison of maximum throughput (averaged over 30 runs) for MQTT, ROS, and \sysname (decentralized).}
      \label{fig:s1-mqtt-ros-xonos}
  \end{minipage}
  \begin{minipage}[b]{\linewidth}
      \centering
      \includegraphics[width=0.7\linewidth]{Figures/Data/Evaluation/key.pdf}
  \end{minipage}\\
  \vspace{1mm}
  \begin{minipage}[b]{\linewidth}
      \centering
      \begin{subfigure}[b]{0.32\linewidth}
          \includegraphics[width=\linewidth]{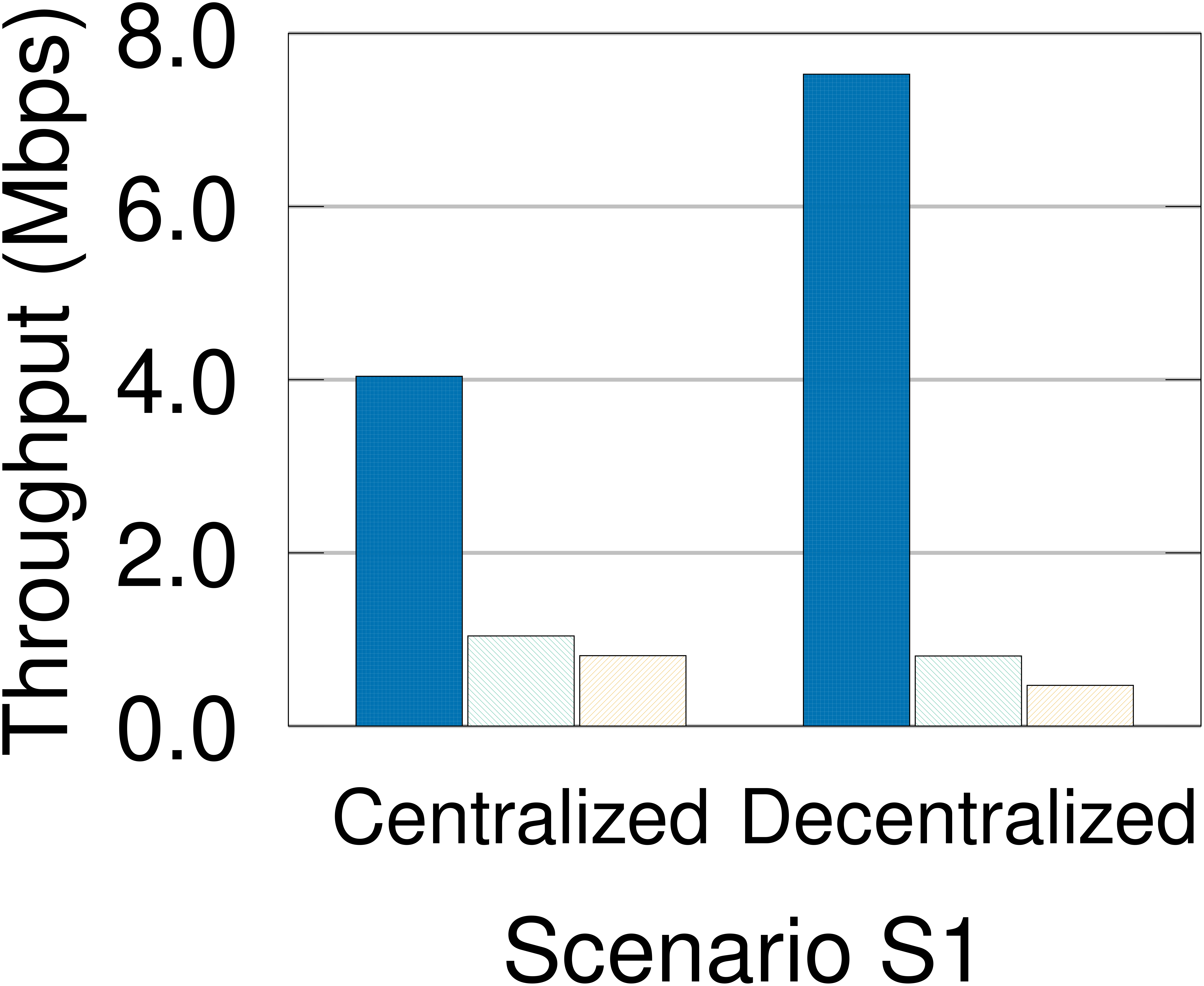}
      \end{subfigure}\hfill
      \begin{subfigure}[b]{0.32\linewidth}
          \includegraphics[width=\linewidth]{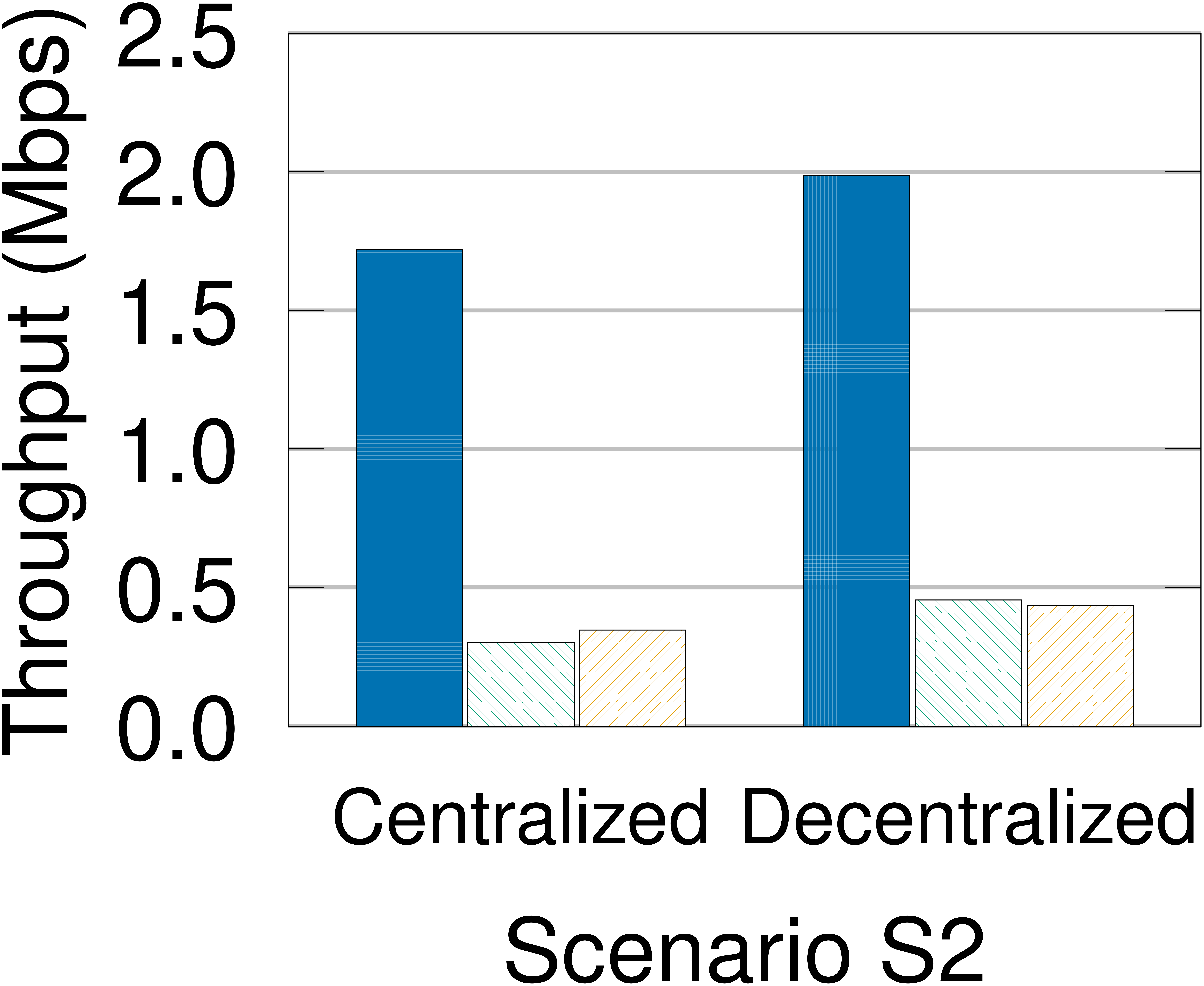}
      \end{subfigure}\hfill
      \begin{subfigure}[b]{0.32\linewidth}
          \includegraphics[width=\linewidth]{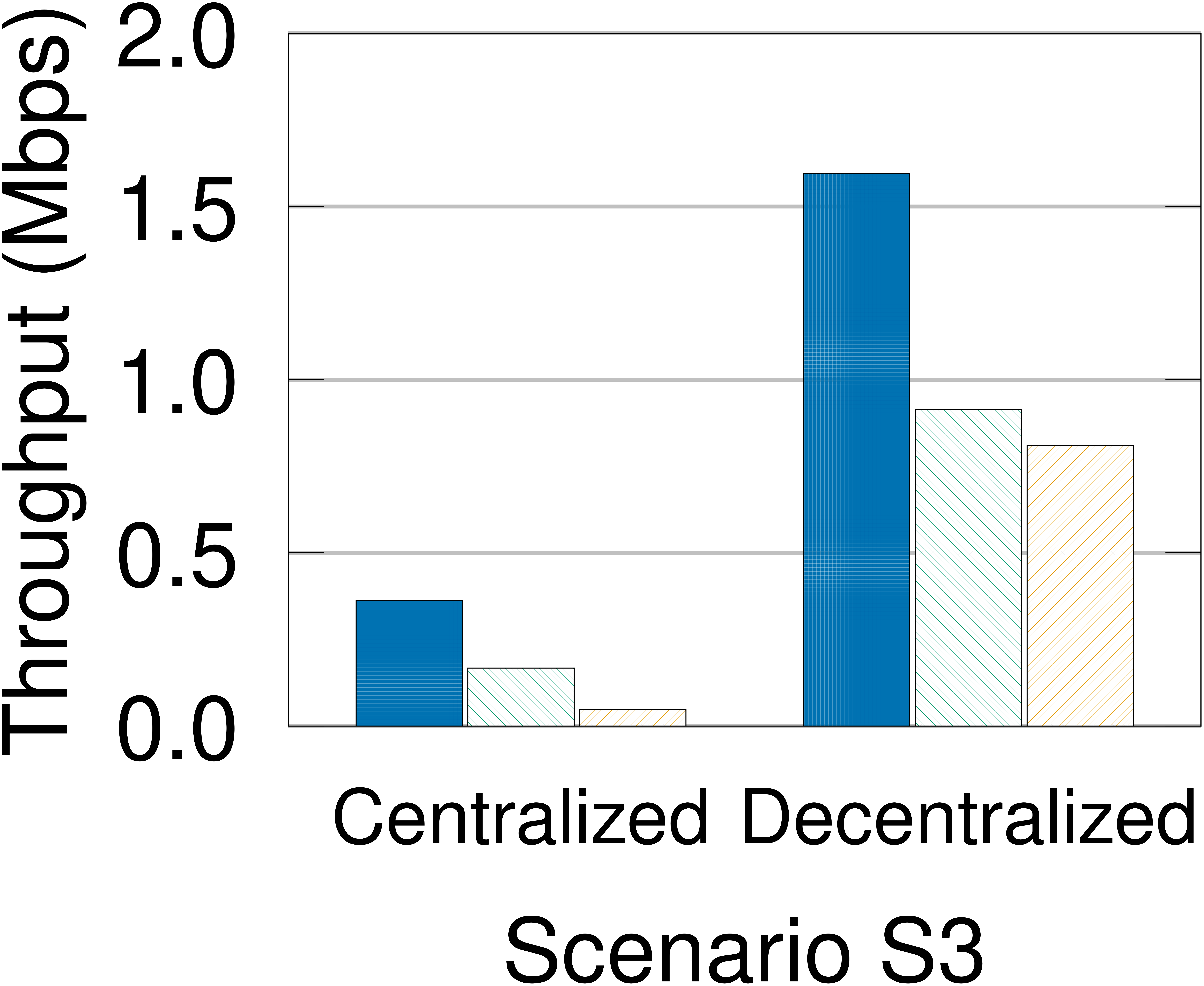}
      \end{subfigure}
      \caption{Maximum throughput (in Mbps) of \sysname under centralized and decentralized coordination.}
      \label{fig:s1-s2-s3-configurations}
    \end{minipage}
    \vspace{-4mm}
  \end{figure}

\subsubsection*{Comparison with baseline} Fig.~\ref{fig:s1-mqtt-ros-xonos} shows
the maximum throughput (in Mbps) under \sysname using decentralized coordination
compared against ROS 2 and MQTT. For this series of comparisons, the data size
is 4 bytes and the STP offset in \sysname is statically calculated for each
benchmark (all in the order of a few milliseconds). 
These measurements are obtained by forcing each benchmark to flood the
network and measuring the total physical time at the
receiver for the total messages received after running the test for 10 seconds
(logical in the case of \sysname and physical for ROS 2 and MQTT, giving an
advantage to the latter). Each test is run at least 30 times. For these
microbenchmarks, we observe that \sysname can provide a higher throughput
than ROS 2. \sysname also has a higher throughput than MQTT for S1 and
S2, but falls short in S3 due to the strict alignment requirement at the Sink.
\subsubsection*{Centralized vs. Decentralized}
While the centralized coordinator imposes a strict global ordering of events, it
does so with additional overhead both in terms of extra control messages and
inserting the RTI as a bottleneck for message communication. Our decentralized
coordination mechanism instead relies on assumptions about the latencies in the
system to ensure correctness.
Fig.~\ref{fig:s1-s2-s3-configurations} shows the comparison between our
coordinators. The results show
that our decentralized coordinator performs better overall, particularly for the
challenging pattern of S3. 
\subsubsection*{Physical vs. Logical} Not all parts of the system need a strict tag ordering in order to
function well. To allow flexibility, we have also added support for physical
connections across federates, as explained in Sec.~\ref{subsec:fedruntime}. This
communication mechanism is the closest match to ROS 2 or MQTT that \sysname
offers.
\begin{table}[t]
	\centering
    \footnotesize
    \caption{Maximum throughput (in Mbps) of physical connections under different coordinations.}
    \begin{tabular}{@{}lccr@{}}
        \toprule
    	\textbf{Coordination}  & \textbf{S1-PC} & \textbf{S2-PC} & \textbf{S3-PC} \\ \midrule
    	Decentralized & 8.16 Mbps & 2.35 Mbps & 2.66 Mbps                         \\
    	Centralized   & 5.81 Mbps &	1.34 Mbps & 1.54 Mbps                         \\ \bottomrule
    \end{tabular}
    \label{tab:physical_connections}
\end{table}
Results are shown in Table~\ref{tab:physical_connections}. We find that physical
connections for these microbenchmarks have up to $7$x higher throughput compared to logical connections.
\begin{table}[t]
	\centering
    \footnotesize
    \caption{Maximum throughput for each supported serialization method, measured over 30 runs, for S1 (in Mbps).}
    \begin{tabular}{@{}lcccr@{}}
    	\toprule
    	\textbf{Serialization}  & \textbf{Avg} & \textbf{Max} & \textbf{Min} &	\textbf{Std} \\ \midrule
    	Native & 4.04 Mbps & 4.27 Mbps & 3.88 Mbps & 0.11 Mbps \\ 
        ROS 2  & 1.19 Mbps & 1.22 Mbps & 1.17 Mbps & 0.2 Mbps \\
        Proto  & 2.50 Mbps & 2.57 Mbps & 2.41 Mbps & 0.05 Mbps \\ \bottomrule
    \end{tabular}
	\label{tab:serialization_overhead}
\end{table}

\begin{table}[t]
	\centering
    \footnotesize
    \caption{Maximum throughput with and without clock synchronization for S1 (in Mbps).}
    \begin{tabular}{@{}lcr@{}}
        \toprule
        \textbf{Coordination} & \textbf{Clock Sync Mode} & \textbf{Max. Throughput} \\ \midrule
        \multirow{3}{*}{Decentralized} & Disabled      & 10.14 Mbps   \\ \cmidrule(l){2-3}
                                       & Startup-only  & 8.16  Mbps   \\ \cmidrule(l){2-3}
                                       & Enabled       & 7.54 Mbps    \\ \midrule
        \multirow{3}{*}{Centralized}   & Disabled      & 6.48 Mbps    \\ \cmidrule(l){2-3}
                                       & Startup-only  & 5.81 Mbps    \\ \cmidrule(l){2-3}
                                       & Enabled       & 5.26 Mbps    \\ \bottomrule
    \end{tabular}
	\label{tab:clock_sync_overhead}
    \vspace{-3mm}
\end{table}

\subsubsection*{Overhead of Serialization}
A benefit of \sysname is that the serialization method used is flexible.
Table.~\ref{tab:serialization_overhead} shows the overhead per message for the
three currently supported serialization methods. We find that while our native
method allows for the highest throughput, Protobufs serialization is also substantially
more performant than ROS 2 serialization. 
\begin{figure}[b]
    \vspace{-3mm}
	\centering
	\includegraphics[width=\linewidth]{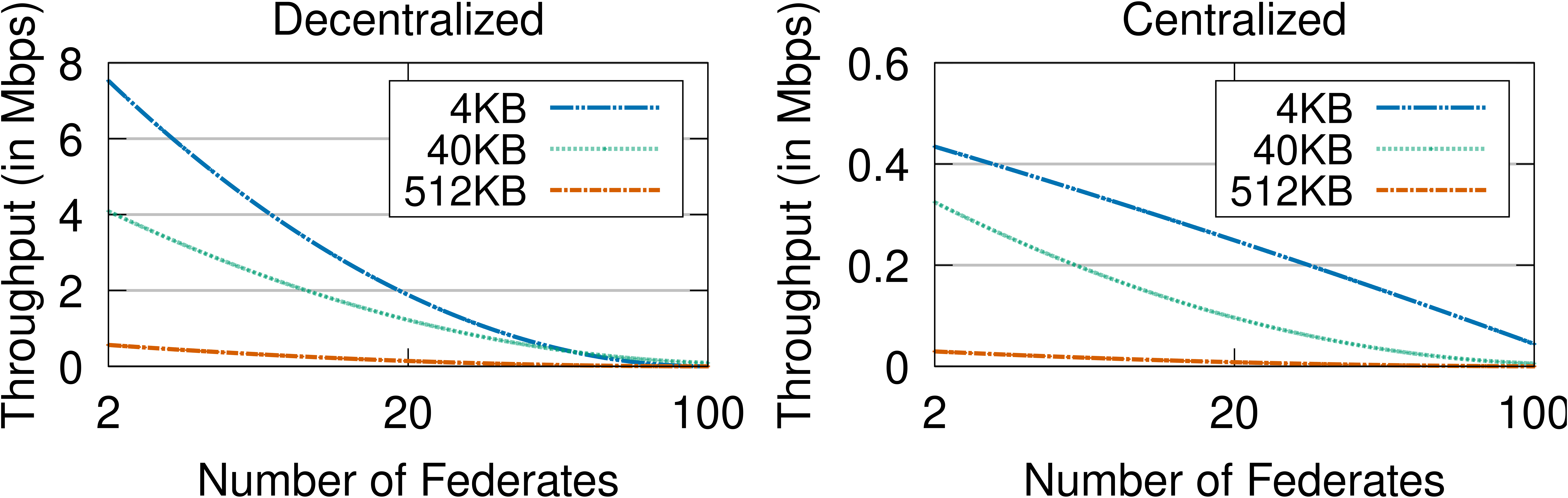}
    \vspace{-2mm}
	\caption{\small{Scalability of throughput (in Mbps) in \sysname for centralized and
	decentralized coordination for S1.}}
	\label{fig:scale}
\end{figure}

\subsubsection*{Overhead of Clock Synchronization}
We measured the throughput of \sysname with and without the built-in clock
synchronization mechanism. 
The results, shown in
Table.~\ref{tab:clock_sync_overhead}, indicate that the impact on throughput of
our built-in clock synchronization is relatively low.
We also found that the clock synchronization error in our method is generally below 10
microseconds.

\subsubsection*{Federation Scalability}
Finally, we tested the scalability of \sysname when number of nodes and message
size is increased. The trends that we observed are depicted in
Fig.~\ref{fig:scale}. We find that throughput roughly scales down linearly with the
number of nodes and message size.

\section{Related Work}\label{sec:related}
%
%
Distributed coordination is a decades-old research topic.
Classic solutions include Chandy and Misra~\cite{ChandyMisra:79:DDE},
Jefferson~\cite{Jefferson:85:TimeWarp}, and HLA~\cite{dahmann1997department}.
PTIDES~\cite{Zou:09:Ptides} offers a decentralized approach, later independently
developed by Google Spanner~\cite{CorbettEtAl:12:Spanner}. Loosely
time-triggered architectures (LTTA)~\cite{baudart2016loosely, TripakisEtAl:08:LTTA} provides a
programming model built upon the strictly synchronous TTA
model~\cite{Kopetz:03:TTA}, generalizing to distributed environments. System-Level LET (SL LET)~\cite{Ernst:18:SLLET} extends Logical Execution Time
(LET)~\cite{Kirsch:12:LET} to distributed settings by relaxing the
synchronization requirements. The reactor model can also be
viewed as a generalization of the LET principle, enabling combinations of
logical execution time with the zero execution time semantics of synchronous
languages while preserving the ability to precisely control the timing of
interactions with the physical environment.

Runtime systems are further developed based on
these coordination schemes. The PRISE project implements a distributed real-time
simulation system based on HLA~\cite{chaudron:2014:RTDistributedSimulationsHLA}.
Ptolemy-HLA framework~\cite{cardoso2018ptolemy}  provides distributed simulation for cyber-physical systems.
TipFrame~\cite{wan2017time} and LETT~\cite{baron2021lett} are examples of
LET-based frameworks. DEAR~\cite{Menard:20:AUTOSAR} is a discrete-event framework for
AUTOSAR~\cite{Furst:09:AUTOSAR}, an emerging industry standard for automotive
software based on reactors.
Our work has overlaps with all of these, as any system solution would, but contributes novel
extensions to distributed coordination mechanisms that preserve determinacy under clearly-stated
assumptions, provide for detection of violations of the assumptions, and provide for handling of the
resulting fault conditions.


\section{Conclusion and Future Work}\label{sec:conclusion} We have shown that
nondeterminism in widely-used pub-sub communication frameworks is potentially
dangerous and can lower the confidence in safety-critical distributed embedded
applications.

We have
extended the \lf coordination language to
support federations while preserving its deterministic semantics, and we
have shown that the cost in performance is manageable.
We give two distributed coordination mechanisms: a centralized one that
emphasizes preserving the program semantics even in the presence of network
failures, and a decentralized one that emphasizes being able to continue to make
forward progress in the presence of network failures. We provide a mechanism for
specifying fault handlers that are triggered by violations of safe-to-process bounds.

\sysname is under active development. There are two main areas for future work: 
(1) Federate recovery, where if a federate fails, the federation would have the capacity 
to continue to operate, and (2) Mutation, where the structure of the federation
would be allowed to change at runtime.

\bibliographystyle{IEEEtran}
\bibliography{Refs}

\begin{thebibliography}{10}
\providecommand{\url}[1]{#1}
\csname url@samestyle\endcsname
\providecommand{\newblock}{\relax}
\providecommand{\bibinfo}[2]{#2}
\providecommand{\BIBentrySTDinterwordspacing}{\spaceskip=0pt\relax}
\providecommand{\BIBentryALTinterwordstretchfactor}{4}
\providecommand{\BIBentryALTinterwordspacing}{\spaceskip=\fontdimen2\font plus
\BIBentryALTinterwordstretchfactor\fontdimen3\font minus
  \fontdimen4\font\relax}
\providecommand{\BIBforeignlanguage}[2]{{%
\expandafter\ifx\csname l@#1\endcsname\relax
\typeout{** WARNING: IEEEtran.bst: No hyphenation pattern has been}%
\typeout{** loaded for the language `#1'. Using the pattern for}%
\typeout{** the default language instead.}%
\else
\language=\csname l@#1\endcsname
\fi
#2}}
\providecommand{\BIBdecl}{\relax}
\BIBdecl

\bibitem{quigley2009ros}
M.~Quigley, K.~Conley, B.~Gerkey, J.~Faust, T.~Foote, J.~Leibs, R.~Wheeler,
  A.~Y. Ng \emph{et~al.}, ``Ros: An open-source robot operating system,'' in
  \emph{ICRA workshop on open source software}, vol.~3.\hskip 1em plus 0.5em
  minus 0.4em\relax Kobe, Japan, 2009, p.~5.

\bibitem{stanford1999mq}
A.~Stanford-Clark and U.~Hunkeler, ``Mq telemetry transport (mqtt),''
  \emph{Online]. http://mqtt. org. Accessed September}, vol.~22, p. 2013, 1999.

\bibitem{light2017mosquitto}
R.~A. Light, ``Mosquitto: server and client implementation of the mqtt
  protocol,'' \emph{Journal of Open Source Software}, vol.~2, no.~13, p. 265,
  2017.

\bibitem{Kato:2018:Autoware}
S.~Kato, S.~Tokunaga, Y.~Maruyama, S.~Maeda, M.~Hirabayashi, Y.~Kitsukawa,
  A.~Monrroy, T.~Ando, Y.~Fujii, and T.~Azumi, ``Autoware on board: Enabling
  autonomous vehicles with embedded systems,'' in \emph{2018 ACM/IEEE 9th
  International Conference on Cyber-Physical Systems (ICCPS)}.\hskip 1em plus
  0.5em minus 0.4em\relax IEEE, 2018, pp. 287--296.

\bibitem{Malavolta:2020:HowToArchitectARobot}
I.~Malavolta, G.~Lewis, B.~Schmerl, P.~Lago, and D.~Garlan, ``How do you
  architect your robots? state of the practice and guidelines for ros-based
  systems,'' in \emph{2020 IEEE/ACM 42nd International Conference on Software
  Engineering: Software Engineering in Practice (ICSE-SEIP)}.\hskip 1em plus
  0.5em minus 0.4em\relax IEEE, 2020, pp. 31--40.

\bibitem{mishra2020use}
B.~Mishra and A.~Kertesz, ``The use of {MQTT} in {M2M} and {IoT} systems: A
  survey,'' \emph{IEEE Access}, vol.~8, pp. 201\,071--201\,086, 2020.

\bibitem{LohstrohEtAl:21:Towards}
M.~Lohstroh, C.~Menard, S.~Bateni, and E.~A. Lee, ``Toward a lingua franca for
  deterministic concurrent systems,'' \emph{ACM Transactions on Embedded
  Computing Systems (TECS), Special Issue on FDL'19}, vol.~20, no.~4, p.
  Article 36, May 2021.

\bibitem{Lohstroh:EECS-2020-235}
\BIBentryALTinterwordspacing
M.~Lohstroh, ``Reactors: A deterministic model of concurrent computation for
  reactive systems,'' Ph.D. dissertation, EECS Department, University of
  California, Berkeley, Dec 2020. [Online]. Available:
  \url{http://www2.eecs.berkeley.edu/Pubs/TechRpts/2020/EECS-2020-235.html}
\BIBentrySTDinterwordspacing

\bibitem{Agha:97:Actors}
G.~A. Agha, ``Abstracting interaction patterns: A programming paradigm for open
  distributed systems,'' in \emph{Formal Methods for Open Object-based
  Distributed Systems, IFIP Transactions}, E.~N. Stefani and J.-B., Eds.\hskip
  1em plus 0.5em minus 0.4em\relax Chapman and Hall, 1997, Conference
  Proceedings.

\bibitem{Kirsch:12:LET}
C.~M. Kirsch and A.~Sokolova, ``The logical execution time paradigm,'' in
  \emph{Advances in Real-Time Systems}.\hskip 1em plus 0.5em minus 0.4em\relax
  Springer, 2012, pp. 103--120.

\bibitem{Benveniste:91:Synchronous}
A.~Benveniste and G.~Berry, ``The synchronous approach to reactive and
  real-time systems,'' \emph{Proceedings of the IEEE}, vol.~79, no.~9, pp.
  1270--1282, 1991.

\bibitem{LeeEtAl:7:DiscreteEvents}
E.~A. Lee, J.~Liu, L.~Muliadi, and H.~Zheng, ``Discrete-event models,'' in
  \emph{System Design, Modeling, and Simulation using {Ptolemy II}},
  C.~Ptolemaeus, Ed.\hskip 1em plus 0.5em minus 0.4em\relax Ptolemy.org, 2014.

\bibitem{Zeigler:1997:DEVS}
B.~P. Zeigler, Y.~Moon, D.~Kim, and G.~Ball, ``The devs environment for
  high-performance modeling and simulation,'' \emph{IEEE Computational Science
  and Engineering}, vol.~4, no.~3, pp. 61--71, 1997.

\bibitem{Liao:97:Scenic}
S.~Liao, S.~Tjiang, and R.~Gupta, ``An efficient implementation of reactivity
  for modeling hardware in the {Scenic} design environment,'' in \emph{Design
  Automation Conference}.\hskip 1em plus 0.5em minus 0.4em\relax ACM, Inc.,
  1997, Conference Proceedings.

\bibitem{LoBelloEtAl:19:TSN}
L.~Lo~Bello and W.~Steiner, ``A perspective on ieee time-sensitive networking
  for industrial communication and automation systems,'' \emph{Proceedings of
  the IEEE}, vol. 107, no.~6, pp. 1094--1120, 2019.

\bibitem{Brewer:12:CAP}
E.~Brewer, ``{CAP} twelve years later: How the "rules" have changed,''
  \emph{IEEE Computer}, vol.~45, no.~2, pp. 23--29, February 2012.

\bibitem{Zhao:07:PTIDES}
Y.~Zhao, E.~A. Lee, and J.~Liu, ``A programming model for time-synchronized
  distributed real-time systems,'' in \emph{Real-Time and Embedded Technology
  and Applications Symposium (RTAS)}.\hskip 1em plus 0.5em minus 0.4em\relax
  IEEE, 2007, Conference Proceedings, pp. 259 -- 268.

\bibitem{CorbettEtAl:12:Spanner}
J.~C. Corbett, J.~Dean, M.~Epstein, A.~Fikes, C.~Frost, J.~Furman, S.~Ghemawat,
  A.~Gubarev, C.~Heiser, P.~Hochschild, W.~Hsieh, S.~Kanthak, E.~Kogan, H.~Li,
  A.~Lloyd, S.~Melnik, D.~Mwaura, D.~Nagle, S.~Quinlan, R.~Rao, L.~Rolig,
  Y.~Saito, M.~Szymaniak, C.~Taylor, R.~Wang, and D.~Woodford, ``Spanner:
  Google's globally-distributed database,'' in \emph{OSDI}, 2012, Conference
  Proceedings.

\bibitem{Lamport:84:TimeStamps}
L.~Lamport, ``Using time instead of timeout for fault-tolerant distributed
  systems,'' \emph{ACM Transactions on Programming Languages and Systems},
  vol.~6, no.~2, pp. 254--280, 1984.

\bibitem{ChandyMisra:79:DDE}
K.~M. Chandy and J.~Misra, ``Distributed simulation: A case study in design and
  verification of distributed programs,'' \emph{IEEE Trans. on Software
  Engineering}, vol.~5, no.~5, pp. 440--452, 1979.

\bibitem{Chandy:86:DDE}
J.~Misra, ``Distributed discrete event simulation,'' \emph{ACM Computing
  Surveys}, vol.~18, no.~1, pp. 39--65, 1986.

\bibitem{dahmann1997department}
J.~S. Dahmann, R.~M. Fujimoto, and R.~M. Weatherly, ``The department of defense
  high level architecture,'' in \emph{Proceedings of the 29th conference on
  Winter simulation}, 1997, pp. 142--149.

\bibitem{Thomas2014}
\BIBentryALTinterwordspacing
D.~Thomas, W.~Woodall, and E.~Fernandez, ``{Next-generation ROS: Building on
  DDS},'' in \emph{ROSCon Chicago 2014}.\hskip 1em plus 0.5em minus 0.4em\relax
  Mountain View, CA: Open Robotics, sep 2014. [Online]. Available:
  \url{https://vimeo.com/106992622}
\BIBentrySTDinterwordspacing

\bibitem{standard2019mqtt}
O.~Standard, ``{MQTT} {Version} 5.0,'' \emph{Retrieved June}, vol.~22, p. 2020,
  2019.

\bibitem{Lohstroh:2019:CyPhy}
M.~Lohstroh, {\'I}.~{\'I}ncer~Romeo, A.~Goens, P.~Derler, J.~Castrillon, E.~A.
  Lee, and A.~Sangiovanni-Vincentelli, ``Reactors: A deterministic model for
  composable reactive systems,'' in \emph{8th International Workshop on
  Model-Based Design of Cyber Physical Systems (CyPhy'19)}, vol. LNCS
  11971.\hskip 1em plus 0.5em minus 0.4em\relax Springer-Verlag, 2019,
  Conference Proceedings, p.~27.

\bibitem{MannaPnueli:93:Verifying}
Z.~Manna and A.~Pnueli, ``Verifying hybrid systems,'' in \emph{Hybrid Systems},
  vol. LNCS 736, 1993, Conference Proceedings, pp. 4--35.

\bibitem{rosenthal2020chaos}
C.~Rosenthal and N.~Jones, \emph{Chaos engineering}.\hskip 1em plus 0.5em minus
  0.4em\relax O'Reilly Media, Incorporated, 2020, vol. 1005.

\bibitem{foote2013tf}
T.~Foote, ``tf: The transform library,'' in \emph{2013 IEEE Conference on
  Technologies for Practical Robot Applications (TePRA)}, 2013, pp. 1--6.

\bibitem{luttgen1999analyzing}
G.~L{\"u}ttgen and V.~Carreno, ``Analyzing mode confusion via model checking,''
  in \emph{International SPIN Workshop on Model Checking of Software}.\hskip
  1em plus 0.5em minus 0.4em\relax Springer, 1999, pp. 120--135.

\bibitem{Zou:09:Ptides}
J.~Zou, S.~Matic, E.~A. Lee, T.~H. Feng, and P.~Derler, ``Execution strategies
  for {Ptides}, a programming model for distributed embedded systems,'' in
  \emph{Real-Time and Embedded Technology and Applications Symposium
  (RTAS)}.\hskip 1em plus 0.5em minus 0.4em\relax IEEE, 2009, Conference
  Proceedings.

\bibitem{Baccelli:92:MaxPlus}
F.~Baccelli, G.~Cohen, G.~J. Olster, and J.~P. Quadrat, \emph{Synchronization
  and Linearity, An Algebra for Discrete Event Systems}.\hskip 1em plus 0.5em
  minus 0.4em\relax New York: Wiley, 1992.

\bibitem{Eidson:06:1588}
J.~C. Eidson, \emph{Measurement, Control, and Communication Using IEEE
  1588}.\hskip 1em plus 0.5em minus 0.4em\relax Springer, 2006.

\bibitem{EidsonStanton:15:Time}
J.~C. Eidson and K.~B. Stanton, ``Timing in cyber-physical systems: the last
  inch problem,'' in \emph{IEEE International Symposium on Precision Clock
  Synchronization for Measurement, Control, and Communication (ISPCS)}.\hskip
  1em plus 0.5em minus 0.4em\relax IEEE, 2015, Conference Proceedings, pp.
  19--24.

\bibitem{GengEtAl:18:ClockSync}
Y.~Geng, S.~Liu, Z.~Yin, A.~Naik, B.~Prabhakar, M.~Rosenblum, and A.~Vahdat,
  ``Exploiting a natural network effect for scalable, fine-grained clock
  synchronization,'' in \emph{15th {USENIX} Symposium on Networked Systems
  Design and Implementation ({NSDI} 18)}, 2018, pp. 81--94.

\bibitem{Black:1998:Dictionary}
P.~E. Black \emph{et~al.}, \emph{Dictionary of algorithms and data
  structures}.\hskip 1em plus 0.5em minus 0.4em\relax Addison-Wesley, 1998.

\bibitem{merkel2014docker}
D.~Merkel \emph{et~al.}, ``Docker: lightweight linux containers for consistent
  development and deployment,'' \emph{Linux journal}, vol. 2014, no. 239, p.~2,
  2014.

\bibitem{LeeLohstroh:2021:TimeForAll}
E.~A. Lee and M.~Lohstroh, ``Time for all programs, not just real-time
  programs,'' in \emph{International Symposium on Leveraging Applications of
  Formal Methods}.\hskip 1em plus 0.5em minus 0.4em\relax Springer, 2021, pp.
  213--232.

\bibitem{Jefferson:85:TimeWarp}
D.~Jefferson, ``Virtual time,'' \emph{ACM Trans. Programming Languages and
  Systems}, vol.~7, no.~3, pp. 404--425, 1985.

\bibitem{baudart2016loosely}
G.~Baudart, A.~Benveniste, and T.~Bourke, ``Loosely time-triggered
  architectures: improvements and comparisons,'' \emph{ACM Transactions on
  Embedded Computing Systems (TECS)}, vol.~15, no.~4, pp. 1--26, 2016.

\bibitem{TripakisEtAl:08:LTTA}
S.~Tripakis, C.~Pinello, A.~Benveniste, S.-V. A., P.~Caspi, and M.~Di~Natale,
  ``Implementing synchronous models on loosely time triggered architectures,''
  \emph{IEEE Transactions on Computers}, vol.~57, no.~10, pp. 1300--1314, 2008.

\bibitem{Kopetz:03:TTA}
H.~Kopetz and G.~Bauer, ``The time-triggered architecture,'' \emph{Proceedings
  of the IEEE}, vol.~91, no.~1, pp. 112--126, 2003.

\bibitem{Ernst:18:SLLET}
R.~Ernst, L.~Ahrendts, and K.-B. Gemlau, ``System level let: Mastering
  cause-effect chains in distributed systems,'' in \emph{IECON 2018-44th Annual
  Conference of the IEEE Industrial Electronics Society}.\hskip 1em plus 0.5em
  minus 0.4em\relax IEEE, 2018, pp. 4084--4089.

\bibitem{chaudron:2014:RTDistributedSimulationsHLA}
J.-B. Chaudron, D.~Saussié, P.~Siron, and M.~Adelantado, ``Real-time
  distributed simulations in an hla framework: Application to aircraft
  simulation,'' \emph{Simulation}, vol.~90, no.~6, pp. 627--643, 2014.

\bibitem{cardoso2018ptolemy}
J.~Cardoso and P.~Siron, ``Ptolemy-hla: A cyber-physical system distributed
  simulation framework,'' in \emph{Principles of Modeling}.\hskip 1em plus
  0.5em minus 0.4em\relax Springer, 2018, pp. 122--142.

\bibitem{wan2017time}
B.~Wan, H.~Luo, K.~Zhou, X.~Li, C.~Wang, X.~Chen, and X.~Zhou, ``A time-aware
  programming framework for constructing predictable real-time systems,'' in
  \emph{2017 IEEE 19th International Conference on High Performance Computing
  and Communications; IEEE 15th International Conference on Smart City; IEEE
  3rd International Conference on Data Science and Systems
  (HPCC/SmartCity/DSS)}.\hskip 1em plus 0.5em minus 0.4em\relax IEEE, 2017, pp.
  578--585.

\bibitem{baron2021lett}
W.~Baron, A.~Arestova, C.~Sippl, K.-S. Hielscher, and R.~German, ``Lett: An
  execution model for distributed real-time systems,'' in \emph{2021 IEEE 94th
  Vehicular Technology Conference (VTC2021-Fall)}.\hskip 1em plus 0.5em minus
  0.4em\relax IEEE, 2021, pp. 1--7.

\bibitem{Menard:20:AUTOSAR}
C.~Menard, A.~Goens, M.~Lohstroh, and J.~Castrillon, ``Achieving derterminism
  in adaptive {AUTOSAR},'' in \emph{Design, Automation and Test in Europe
  ({DATE} 20)}, Grenoble, France, March 2020.

\bibitem{Furst:09:AUTOSAR}
S.~F{\"u}rst, J.~M{\"o}ssinger, S.~Bunzel, T.~Weber, F.~Kirschke-Biller,
  P.~Heitk{\"a}mper, G.~Kinkelin, K.~Nishikawa, and K.~Lange, ``Autosar--a
  worldwide standard is on the road,'' in \emph{14th International VDI Congress
  Electronic Systems for Vehicles, Baden-Baden}, vol.~62, 2009, p.~5.

\end{thebibliography}

\end{document}